\documentclass[a4paper,11pt]{article}
\pdfoutput=1 
\usepackage{jcappub} 

\usepackage{qubic}
\usepackage{newtxtext,newtxmath}
\usepackage{comment}
\usepackage{mathtools}

\title{QUBIC VIII: Optical design and performance}
\author[1]{C.~O'Sullivan}
\author[2,3]{M.~De~Petris}
\author[2]{G.~Amico}
\author[2,3]{E.S.~Battistelli}
\author[2,3]{P.~de~Bernardis}
\author[1]{D.~Burke}
\author[2]{D.~Buzi}
\author[4]{C.~Chapron}
\author[5]{L.~Conversi}
\author[2,3]{G.~D'Alessandro}
% ...

\affiliation[1]{National University of Ireland, Maynooth, Ireland}
\affiliation[2]{Universit\a`{a} di Roma - La Sapienza, Roma, Italy}
\affiliation[3]{INFN sezione di Roma, 00185 Roma, Italy}
\affiliation[4]{Universit\'e de Paris, CNRS, Astroparticule et Cosmologie, F-75006 Paris, France}
\affiliation[5]{ESA/ESRIN, Frascati (RM), Italy}
% ...

\date{\today}
\emailAdd{creidhe.osullivan@mu.ie}
\emailAdd{marco.depetris@roma1.infn.it}

\abstract{The Q and U Bolometric Interferometer for Cosmology (QUBIC) is a ground-based experiment that aims to detect B-mode polarization anisotropies \cite{Zaldarriaga1997} in the CMB at angular scales around the $\ell\simeq$100 recombination peak. Systematic errors make ground-based observations of B modes at millimetre wavelengths very challenging and QUBIC mitigates these problems in a somewhat complementary way to other existing or planned experiments using the novel technique of bolometric interferometry. This technique takes advantage of the sensitivity of an imager and the systematic error control of an interferometer. A cold reflective optical combiner superimposes the re-emitted beams from 400 aperture feedhorns on two focal planes. A shielding system composed of a fixed groundshield, and a forebaffle that moves with the instrument, limits the impact of local contaminants.
The modelling, design, manufacturing and preliminary measurements of the optical components are described in this paper.}


\begin{document}
\maketitle

\tableofcontents

\section{Introduction}{\label{sec:intro}}


The Q and U Bolometric Interferometer for Cosmology (QUBIC) is a ground-based telescope designed to detect B-mode polarization anisotropies in the Cosmic Microwave Background (CMB).  It will observe at 150~GHz, in the first instance, and sample the range of multipoles $l\approx30-200$ in the sky. A 220-GHz band will be added later to improve astrophysical foreground subtraction. QUBIC will be installed in Argentina, near the city of San Antonio de los Cobres, at the Alto Chorrillos site (4869 m a.s.l.), Salta Province. A detailed overview of the QUBIC project and its scientific goals are given in a companion paper \cite{2020.QUBIC.PAPER1} and in Battistelli \textit{et al.} \cite{Battistelli2011}. Its main parameters are summarized in table~\ref{tab:qubic_params}. 

\begin{table}[t]
    \renewcommand{\arraystretch}{1.}
    \begin{center}
        \caption{\label{tab:qubic_params}QUBIC main parameters (from Hamilton \textit{et al.} \cite{2020.QUBIC.PAPER1})}
        \begin{tabular}{p{5cm}  p{5.5cm}}
            \hline
            Parameter & Full Instrument value \\
            \hline
            \hline
            Frequency channels \dotfill  & 150 GHz \& 220 GHz\\
            Frequency range 150 GHz \dotfill  &[131-169] GHz\\
            Frequency range 220 GHz \dotfill  &[192.5-247.5] GHz\\
            Window Aperture [m]\dotfill  & 0.56 \\
            Focal plane temp. [mK]\dotfill  &300\\
            Sky Coverage\dotfill  &1.5\%\\
            FWHM [degrees]\dotfill &0.39 (150 GHz), 0.27 (220 GHz)\\
        \end{tabular}
        
    \end{center}
\end{table}

\begin{figure}
\centering
\includegraphics[ width = 1.0 \hsize]{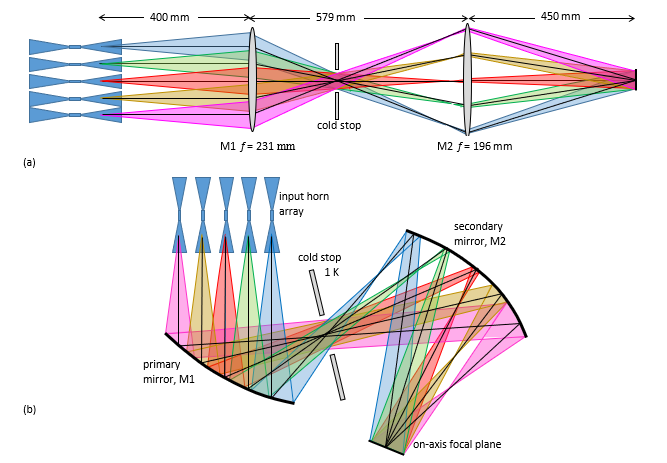}
   \caption{(a) An ideal imager superimposing the beams from an input array of horns. Here the imager is represented as a pair of lenses, in practice QUBIC uses mirrors (b).  (Ray tracing in \swname{Zemax} software (https://www.zemax.com/) was used to produce the beams in this diagram.) The optical system in (a) is the on-axis equivalent of the QUBIC combiner, with a final effective focal length equal to 300 mm. Beams from five locations in the aperture array were selected as examples. (Adapted from O'Sullivan \textit{et al.}~\cite{OSullivanSPIE2018Optics}.)}
   \label{figidealrealray}
\end{figure}

QUBIC operates as a Fizeau interferometer; the sky radiation is collected from an array of back-to-back feedhorns at the entrance aperture of the instrument and their exit beams are combined.  For QUBIC, the beam combiner is an optical imager that superimposes the beams on a focal plane (figure~\ref{figidealrealray}).  The beams from any pair of horns produce a fringe pattern on this focal plane and, for perfect imaging, equivalent baselines produce identical fringe patterns, a property we take advantage of in a `self-calibration' technique \cite{BigotSazy2013}.  The fringe pattern image is sampled by an array of bolometers, hence the term \textit{bolometric interferometry}.  The complex fringe visibility could be measured from this image of fringes but QUBIC will be used as a synthetic imager, opening a mechanical switch in all the horns and thereby observing the fringes from all baselines simultaneously.  The combined fringe pattern is simply an image of the sky convolved with the synthesized beam of the instrument. This synthesized beam is the Fourier transform of the aperture field and is largely determined by the location of the horns in the input array (the maximum baseline determines its FWHM).  This is similar to the technique of aperture-masked interferometry in optical astronomy; here the array of horn apertures acts as the aperture mask.  The field-of-view of the instrument is determined by the beam pattern of the individual horns.

 In QUBIC, the full optical chain is completely contained within a cryostat. The signal from the sky enters through a window and series of filters and is modulated by a rotating half-wave plate (HWP).  If the incoming radiation is written in terms of its electric field in orthogonal directions, $E_{x}$ and $E_{y}$, then after the HWP  the signal in those directions will be

\begin{equation}
S_{\mathrm{HWP}} = \begin{pmatrix}
E_{x}\mbox{cos}2\phi(t)+E_{y}\mbox{sin}2\phi(t)\\
E_{x}\mbox{sin}2\phi(t)-E_{y}\mbox{cos}2\phi(t)
\end{pmatrix}
\end{equation}

\noindent
where $\phi(t)$ is the time-varying angle of rotation of the HWP.  A polarizing grid then removes the signal in one direction giving

\begin{equation}
S_{\mathrm{HWP}} = \begin{pmatrix}
E_{x}\mbox{cos}2\phi(t)+E_{y}\mbox{sin}2\phi(t)\\
0
\end{pmatrix}
\end{equation}

\noindent
or, in terms of the Stokes’ parameters, $S=I+Q\mbox{cos}4\phi(t)+U\mbox{sin}4\phi(t)$.  It is this signal that is convolved with the synthesized beam of the instrument. The modulation of the signal as a function of HWP angle $\phi(t)$ allows $I$, $Q$ and $U$ to be reconstructed.  A dichroic filter may be used to split the signal into two frequency bands that can then be imaged simultaneously on orthogonal focal planes (as illustrated in figure~\ref{figCAD}).

We propose this technique as an alternative to existing B-mode instruments with optics based on refractive or reflective telescopes (see e.g. Ali \textit{et al.}~~\cite{Ali2020}, St. Germaine \textit{et al.}~\cite{StGermaine2020}). The first experiment to employ such an observational strategy at millimetre wavelengths was the Millimeter-Wave Bolometric Interferometer (MBI) \cite{Timbie2006},\cite{MBI_2008}. MBI was a prototype Fizeau interferometer with a limited number of horns, only four, and it used a Cassegrain telescope as an optical combiner. MBI made observations in 2008 and 2009.

In this paper we report the design, modelling and testing of the optics of the QUBIC instrument, concentrating on the optical beam combiner and the shielding. Following a description of the instrument, the filters, HWP and polarizer are discussed in section~\ref{sec:HWP}. In section~\ref{sec:horns} we describe the feedhorn array that collects radiation from the sky and re-emits it in the combiner. Section~\ref{sec:combiner} deals with the challenging design of the optical combiner and with its detailed modelling using physical optics (PO). A demonstrator instrument is described in section~\ref{subsec:TD} where we also discuss the manufacture of its mirrors and the optical measurements taken.  Finally, in section~\ref{sec:GSFB}, we describe the shielding system that is composed of a forebaffle and a groundshield, designed to reduce contamination from unwanted sources.

\section{The QUBIC instrument}{\label{subsec:intro_instrument}}

\subsection{Optical requirements} {\label{subsubsec:intro_requirements}}

The optical combiner must act as an imager and have an unobstructed aperture that accommodates an array of 400 horns with 14-mm spacing.  The overall focal length of the combiner must be sufficiently long so that the narrowest fringes (from $l\approx30$ features on the sky) are sampled by 3-mm bolometers and not so long that the focal plane would require more than $\sim1000$ bolometers to image the largest fringes (from $l\approx200$ features). The field-of-view of the instrument, determined by the FWHM of the sky-facing horns, is $12.9^{\circ}$ at 150~GHz.  It is to operate in two frequency bands centred at 150 and 220~GHz.  An overall loss in efficiency of up to 10\% caused by any truncation or aberration of the beams re-emitted within the combiner was considered tolerable. A suitable location for a cold-shield aperture would help reduce straylight on the bare bolometer array. The full combiner must fit into a cryostat of $1~\mbox{m}^{-3}$. We designed an off-axis dual reflector that satisfied these criteria and is described further in section~\ref{subsec:combinerdesign}.

\subsection{Instrument overview}{\label{subsubsec:intro_overview}}

\begin{figure}
\centering
\includegraphics[ width = 0.65 \hsize ]{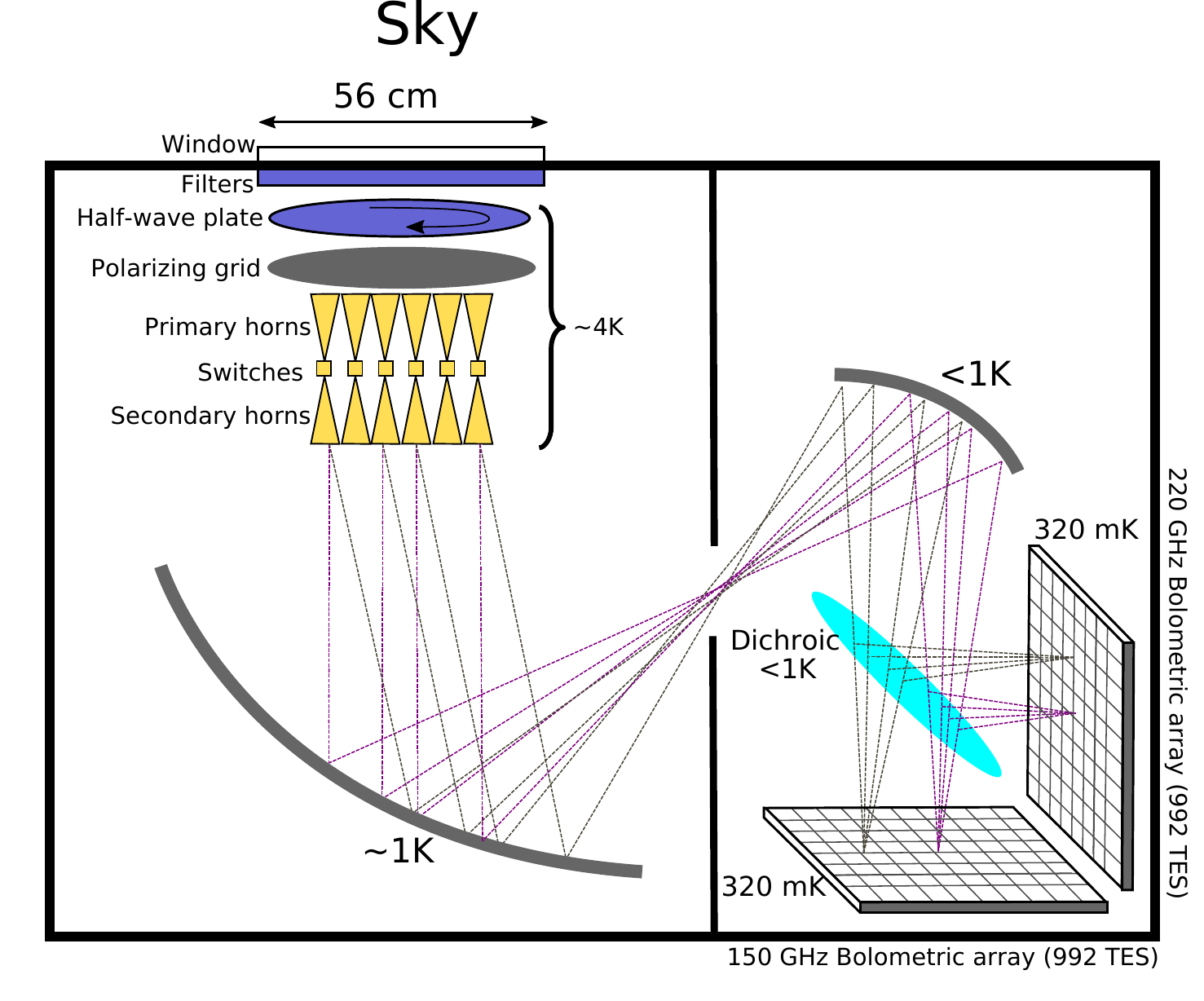}
\includegraphics[ width = 0.75 \hsize ]{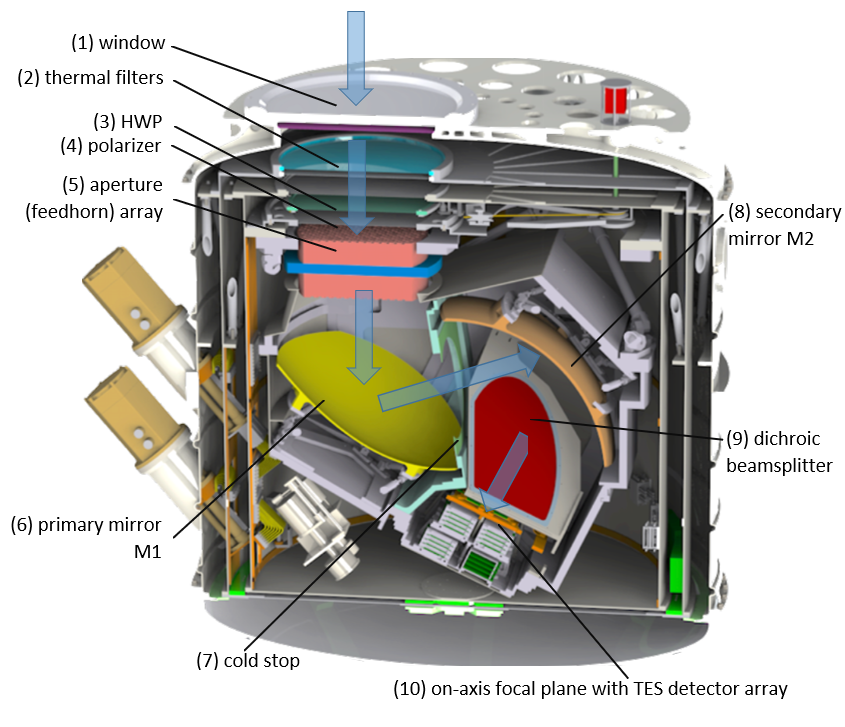}
   \caption{(top) Schematic and (bottom) CAD drawing of the QUBIC optical and detection chain.}
   \label{figCAD}
\end{figure}

A sketch of the QUBIC instrument is shown in figure~\ref{figCAD}. The full optical chain is housed in the large QUBIC cryostat whose main instrument volume is cooled to 4~K using two 1-W pulse tube cryocoolers.  The sky signal enters the cryostat through (1) a 560-mm diameter window made from high-density polyethylene (HDPE).  After the window there is (2) a series of thermal (infrared (IR)) and low-pass blocking filters - far-infrared (FIR) blocking and spectral-band defining filters are located at thermal stage apertures further along the optical chain. The polarization of the incoming signal is modulated by (3) a rotating HWP before a single polarization is selected by (4) the polarizer. The signal is collected and then re-emitted into the combiner by (5) a circular array of 400 back-to-back corrugated feedhorns each of which can be opened and closed by means of a mechanical shutter.  The horn beams are then reflected by (6) a primary and (8) secondary mirror onto (10) the 320-mK focal plane. (7) A cold stop is inserted  between the primary and secondary mirrors to minimize stray light. (9) A dichroic beamsplitter will be used to reflect the 220-GHz band onto an orthogonal plane.

This QUBIC instrument (that we refer to as the full instrument (FI)) could constrain the tensor-to-scalar ratio down to $\sigma(r)=0.015$ after a three-year survey~\cite{QUBICDesignReport2016}. If successful, it is hoped to add further similar QUBIC modules in the future.
A QUBIC technological demonstrator (TD) has already been built to validate the QUBIC design and test it electrically, thermally and optically. The full results of the TD calibration campaign are described in detail by Torchinsky \textit{et al.}~\cite{2020.QUBIC.PAPER3}. The simulations shown in this paper were mainly carried out at 150~GHz, the centre of the TD frequency band for which we have measurements.

\section{Filters, HWP and polarizer}{\label{sec:HWP}}
\subsection{Filters}{\label{subsec:filters}}
The full optical chain is housed in the QUBIC cryostat \cite{2020.QUBIC.PAPER5} which consists of several thermal stages. The main instrument volume is cooled to 4~K and the surrounding radiation shield to 40~K.  The mirrors are mounted in a 1-K box and the TES detectors in a 320-mK stage.
The observed radiation is spectrally selected into the two main bands by the combination of a filter chain and feedhorn geometry (high-pass). The optical selective components are distributed at different thermal stages depending on the expected radiation loading.

The optical and thermal environment of the instrument is controlled (in part) by Cardiff metal mesh filters. A combination of thin IR thermal blocking filters and multi-layer sub-mm low pass edge filters (LPE) provides the blocking of unwanted radiation reaching the 320-mK (detector) stage.
Successive thermal blocking filters are placed between 300~K and 4~K to reflect the IR radiation emitted by these stages and to protect the thermal environment at the 1-K and 320-mK stages. These filters are very thin and have extremely high (>99\%) in band transmission.
The LPE filters are placed at 4~K, 1~K and 320~mK and are of standard Cardiff multi-layered capacitive mesh design \cite{AdeSPIE2006}. These devices set a range of well-defined cut-off frequencies in the FIR regime. Typically each filter consists of between 6 and 12 meshes, with $\lambda$/4 spacing, which are embedded in a low-loss polypropylene dielectric. Their cut-off frequencies are chosen in order to be close to the operational band edges, to suppress successive out-of-band leaks, and the designs are optimized for high in-band transmission. Torchinsky \textit{et al.} \cite{2020.QUBIC.PAPER3} discuss measurements of the filter bandpasses. A dichroic filter is under development for the FI but operation at high angles of incidence will be challenging. An alternative for QUBIC is to exploit its spectral-imaging capability within one wide band for foreground subtraction. This is described in an accompanying paper \cite{2020.QUBIC.PAPER2}.

\subsection{HWP and polarizer}{\label{subsec:HWP}}
QUBIC uses a stepped rotating HWP to modulate the incoming polarization before a single polarization is selected by a polarizing grid. The QUBIC HWP is based on embedded metal-mesh filter technology as described by Pisano \textit{et al.} \cite{2012.HWP}.  It has 12 anisotropic mesh grids embedded in polypropylene with an overall thickness of 4.1 mm and a diameter of 370 mm. The design, modelling and measurement of a similar large-badwidth metamaterial-based HWP has been detailed elsewhere \cite{2020.HWP}.  A smaller 180-mm diameter version has been manufactured for testing in the TD (section ~\ref{subsec:TD} and Torchinsky \textit{et al.}~\cite{2020.QUBIC.PAPER3}).  The HWP can be stepped between 7 positions, separated by $15^{\circ}$, ranging from $0^{\circ}$ to $90^{\circ}$, to a precision of $< 0.2^{\circ}$ at cryogenic temperatures.  The HWP rotator mechanism is described in detail by D’Alessandro \textit{et al.} \cite{2020.QUBIC.PAPER6}. The grid is a 10-µm period lithographically etched copper wire device. The ‘wires’ are 5 µm wide and 400 nm thick; the substrate is 1.9 µm Mylar. The polarizer and HWP were both manufactured in Cardiff \cite{2014PisanoHWP}. Calibration tests described by Torchinsky \textit{et al.} \cite{2020.QUBIC.PAPER3} show greater than 99.6\% cross-polarisation rejection by the HWP and grid combination measured on the TD focal plane.

\section{Feedhorn array}{\label{sec:horns}}
\subsection{Back-to-back feedhorns}
\begin{figure}
\centering
\includegraphics[ width = 0.7 \hsize ]{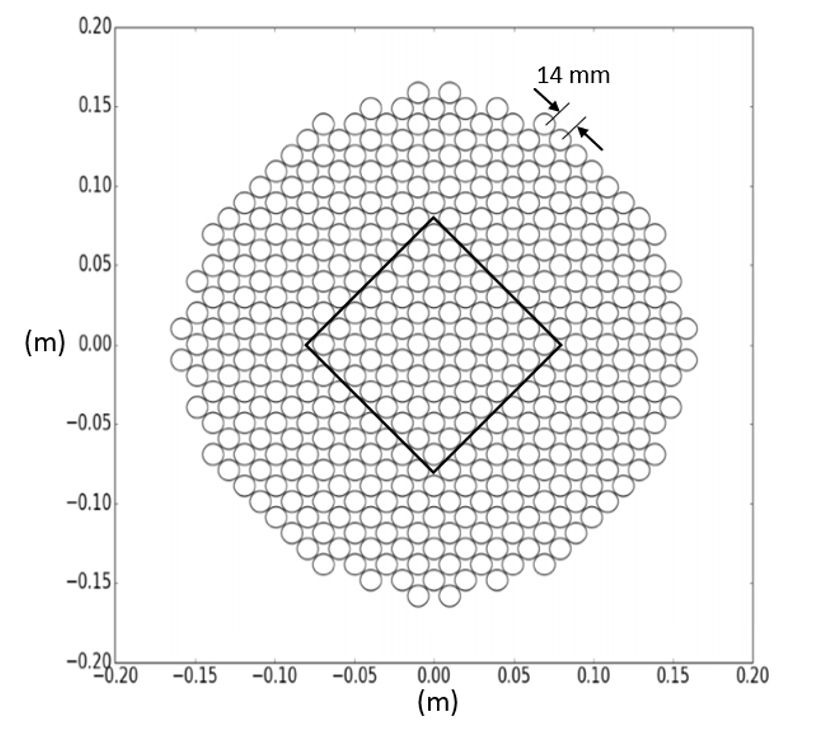}
\includegraphics[ width = 0.29 \hsize ]{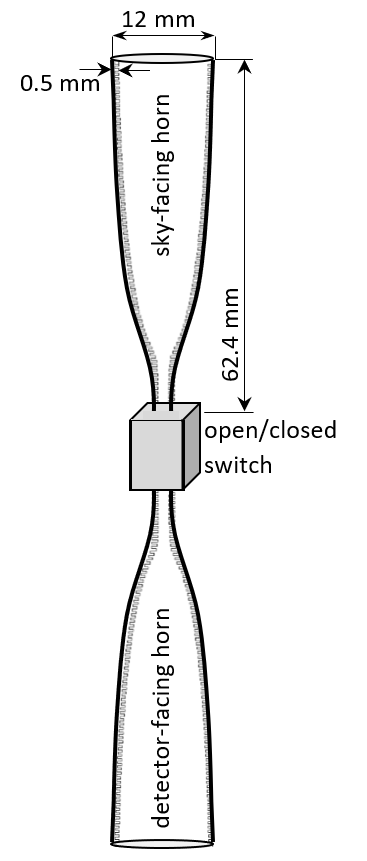}
   \caption{(left) Schematic diagram of the layout of the 400 feedhorns in the input aperture array. The selection of 64 feedhorns used in the technological demonstrator (section~\ref{subsec:TD}) is also indicated. (right) Geometry of a single back-to-back pair of horns \cite{2020.QUBIC.PAPER7}.}
   \label{hornlayout}
\end{figure}

{\label{subsec:b2bfeedhorns}}
The QUBIC input aperture consists of an array of 400 back-to-back conical corrugated horn pairs, laid out in a circular pattern as shown in figure~\ref{hornlayout}. The array is kept cold inside a cryostat and located behind the cryostat window that holds the vacuum, a thermal filter, a HWP and a polarizer. One feedhorn of each pair collects radiation from the sky and the other re-emits the signal onto the beam combiner. A shutter is placed in the waveguide section joining each back-to-back horn pair so signals can be switched on and off in the calibration procedure. The sky-facing and detector-facing feedhorns are identical and were designed to have a FWHM of 12.9$^{\circ}$ at 150~GHz.  The feedhorn design was optimized to operate in two frequency bands (i.e. 130–170~GHz and 190–240~GHz) simultaneously. A detailed discussion of the feedhorn design and performance can be found in Cavaliere \textit{et al.} \cite{2020.QUBIC.PAPER7}.

\subsection{Feedhorn modelling}
\begin{figure}
\centering
\includegraphics[ width = 0.65 \hsize ]{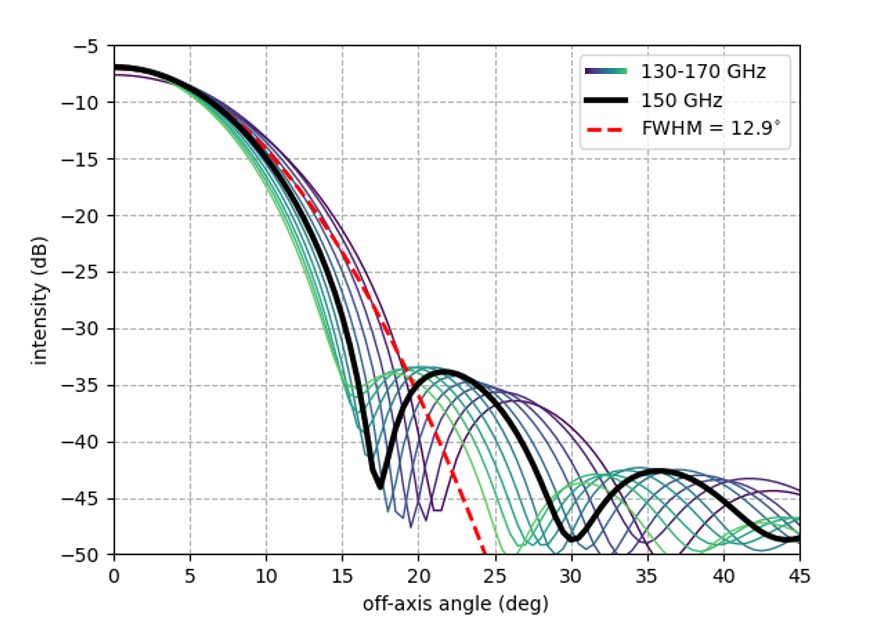}
   \caption{Farfield beam patterns calculated at 4-GHz intervals across the single-moded 150-GHz band. A Gaussian beam with a FWHM of 12.9$^{\circ}$ is shown for comparison.}
   \label{figbeam150}
\end{figure}

\begin{figure}
\centering
\includegraphics[ width = 0.65 \hsize ]{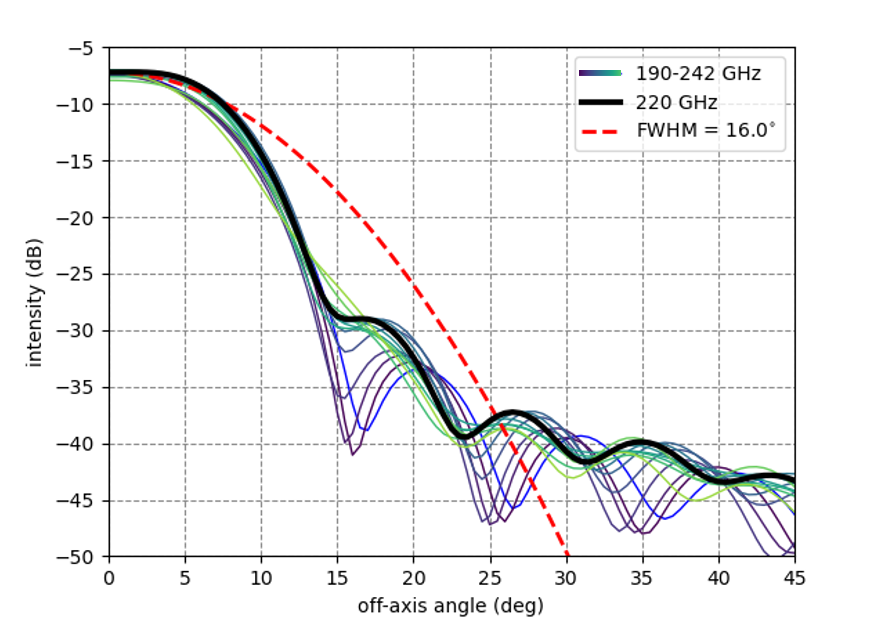}
   \caption{Farfield beam patterns calculated at 4-GHz intervals across the multi-moded 220-GHz band. A Gaussian beam with a FWHM of 16.0$^{\circ}$ (the FWHM at 220~GHz) is shown for comparison. The more top-hat shape of the horn beams is clear.}
   \label{figbeam220}
\end{figure}

{\label{subsec:feedmodelling}}
We use a rigorous electromagnetic mode-matching technique \cite{Clarricoats1984} to model the exact profile of the feedhorns in the input array. This technique regards the corrugated structure as a sequence of smooth-walled cylindrical wave-guide sections, each of which can support a set of TE$_{nm}$ and TM$_{nm}$ modes. At each corrugation there is a sudden change in the radius of the cylindrical guide and this change results in a scattering of power into backward propagating reflected modes and forward propagating transmitted  modes. The scattering between modes is determined by matching the total transverse fields at each junction so that total complex power is conserved.  We find that at least 40 such TE and TM modes are required to adequately describe the QUBIC beam out to the first few sidelobes.  Sets of the TE and TM modes maintain a fixed mutual phase relationship and it is these sets that are the smaller number of hybrid HE and EH modes that are usually referred to when discussing corrugated horns. Tables \ref{tab:150modes} and \ref{tab:220modes} show the power in modes of different azimuthal order at various frequencies across the band.  In the lower frequency band the horns are single-moded (HE$_{11}$) whereas in the upper band three modes can propagate (HE$_{11}$, E$_{02}($=TM$_{02}$), EH$_{21}$) \cite{ScullyPhD}. The HE$_{11}$ and EH$_{21}$ modes also have orthogonal modes that are transmitted. In the lower band the beam has a FWHM of 12.9$^{\circ}$ at 150~GHz and scales with frequency as expected (figure~\ref{figbeam150}).  It can be well approximated by a Gaussian beam (amplitude) of waist radius \cite{Goldsmith1998}

\begin{equation}
w_{\mathrm{o}} = \frac{ \sqrt{2\ln(2)} \lambda}{\pi\theta_{\mathrm{FWHM}}}
\end{equation}

\noindent
where $\theta_{\mathrm{FWHM}}$ is the intensity FWHM angle in the farfield. However, for detailed modelling, we use the full description of the horn field.  In the upper band the main beam is more top-hat in shape and varies less over the band due to its multi-mode nature (figure~\ref{figbeam220}).  When modelling the QUBIC optics, described next, we propagate each mode separately and then add the resulting intensities on the plane of interest.  Our mode-matching software also allows us to calculate return loss (S$_{11}$) if required \cite{EuCAP2020},\cite{2020.QUBIC.PAPER7}.

\begin{table}
\caption{\label{tab:150modes} Power in the hybrid modes present at the output of the QUBIC feedhorn, assuming unit input power for each mode, in the 150~GHz band.}
\centering
\renewcommand{\arraystretch}{1.0}
\begin{tabular}{c|c|c|c} 
 \hline
 frequency & \multicolumn{3}{c}{azimuthal order}\\ 

 (GHz) & 0 & 1 & 2 \\
 \hline\hline
 130 &   & 0.8894 &  \\ 
 135 &   & 0.9851 &   \\ 
 140 &   & 0.9962 &   \\ 
 145 &   & 0.9978 &   \\ 
 150 &   & 0.9947 &   \\ 
 155 &   & 0.9923 &   \\ 
 160 &   & 0.9980 &   \\ 
 165 & $10^{-12}$ & 0.9985  & 0.0123\\ 
 170 & $10^{-9}$ & 0.9965  & $10^{-6}$\\ 
\hline
\end{tabular}
\end{table}

\begin{table}
\caption{\label{tab:220modes} Power in the hybrid modes present at the output of the QUBIC feedhorn, assuming unit input power for each mode, in the 220~GHz band.}
\centering
\renewcommand{\arraystretch}{1.0}%
\begin{tabular}{ c|c|c|c|c|c } 
 \hline
 frequency & \multicolumn{5}{c}{azimuthal order}\\
 (GHz) & 0 & \multicolumn{2}{c}{1} & 2 & 3 \\
 \hline\hline
 190 & 0.9928 & 0.9996 &   & 0.60$\times10^{-7}$ & \\
 200 & 0.9866 & 0.9993 &   & 0.29$\times10^{-3}$ & \\ 
 210 & 0.9939 & 0.9998 &   & 0.9438 & $10^{-15}$\\ 
 220 & 0.9795 & 0.9995 &   & 0.9868 &   \\ 
 230 & 0.9671 & 0.9933 &   & 0.9989 & \\ 
 240 & 0.8807 & 0.9983 &   & 0.9856 & \\ 
 245 & 0.8807 & 0.7884 & 0.0010  & 0.9852 &  \\ 
 248 & 0.4217 & 0.9255 & 0.3312  & 0.9087 & \\ 
 250 & 0.0017 & 0.5540 & 0.1972  & 0.9393 & \\ 
\hline
\end{tabular}
\end{table}

\subsection{Manufacturing and measurements}{\label{subsec:manuHorn}}

The feedhorn array was manufactured using a platelet technique where holes were chemically etched into 176 0.3-mm thick aluminium platelets corresponding to the 88 corrugations of the horn design. We used our mode-matching software to carry out a tolerance analysis on the hole radius \cite{EuCAP2020},\cite{2020.QUBIC.PAPER7}. The array was made in four quarters that are held together by two 3-mm thick flanges (top and bottom) milled from solid aluminium. The full block is mechanically clamped using ERGAL screws. A movable shutter placed in the middle of each back-to-back pair acts as a switch so that individual baselines can be selected during self-calibration. The feedhorn and switch array is described in detail in a companion paper \cite{2020.QUBIC.PAPER7}.

\section{Optical combiner}{\label{sec:combiner}}

\subsection{Ideal combiner}{\label{subsec:combinerideal}}
The combiner is used to superimpose, on the focal plane, the beams from the re-emitting (detector-facing) horns. It therefore must be a focusing system and, in the absence of truncation and aberrations, the focal plane pattern is the Fourier Transform of the feedhorn array pattern.  For an on-axis point source and a single baseline (pair of open horn switches separated by a distance $\Delta x$), the pattern is a series of fringes with a spatial frequency that depends on the baseline length, $\Delta x / \lambda$, where $\lambda$ is the wavelength, and at an angle that depends on the baseline orientation. Equivalent baselines will result in identical fringe patterns.  When all the horn switches are open, as will be the case when QUBIC is observing, the focal plane pattern consists of an array of peaks whose separation, $ \lambda / \Delta h$, depends on the horn separation, $ \Delta h $, and whose width, $\sim \lambda / ((P-1) \Delta h)$, depends on the extent of the horn array, $(P-1) \Delta h$, where $P$ is the number of feedhorns along one side of the array.  This pattern is multiplied by an envelope which is the Fourier transform of the re-emitting feedhorn aperture pattern (i.e. its far-field beam pattern). This is the standard result for a diffraction grating in the Fraunhofer limit with a Gaussian rather than the usual sinc envelope.  For an off-axis source the whole pattern is further multiplied by a factor which accounts for the coupling of the source to each horn antenna.  This factor is simply the beam pattern of the sky-facing feedhorn at the source angle. We refer to the focal plane pattern resulting from a far-field point source in a direction $\textbf{n}$ as the point-spread-function ($\mbox{PSF}_{\textbf{n}}$):  

\begin{equation}
\mbox{PSF}_{\textbf{n}}(\textbf{r})=B_{\mathrm{sky}}(\textbf{n})B_{\mathrm{det}}(\textbf{r})
\frac{\mbox{sin}^2[P \pi \frac{\Delta h}{\lambda} (\frac{x}{f}-n_x)]}{\mbox{sin}^2[\pi \frac{\Delta h}{\lambda} (\frac{x}{f}-n_x)]} \frac{\mbox{sin}^2[P \pi \frac{\Delta h}{\lambda} (\frac{y}{f}-n_y)]}{\mbox{sin}^2[\pi \frac{\Delta h}{\lambda} (\frac{y}{f}-n_y)]}
\end{equation}

\noindent
where $\textbf{r}=(x,y)$ is position on the focal plane, $\textbf{n}=(n_x,n_y)$ is the off-axis direction of the point source, subscripts $x$ and $y$ refer to projection along the axes of the focal plane, $B_{\mathrm{sky}}$ and $B_{\mathrm{det}}$ are the sky-facing and detector-facing beam pattern intensities, $P$ is the number of horns along a side of a square array and $f$ is the focal length of the combiner.

\begin{figure}
\centering
\includegraphics[ width = 0.6 \hsize ]{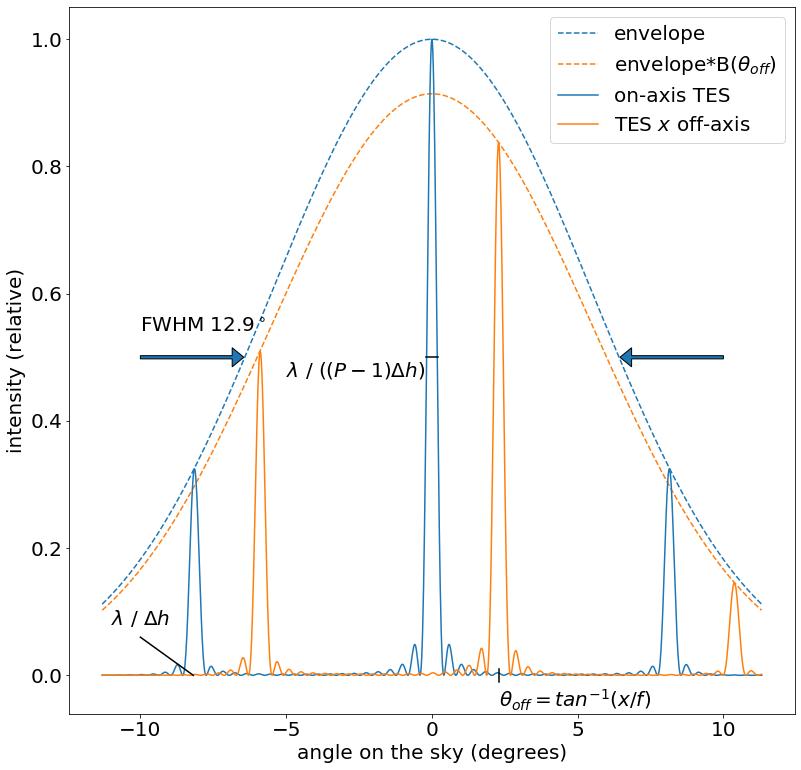}
   \caption{Ideal synthesized beams $B_{{(0,0)}}({\theta})$ (blue) and $B_{{(x,0)}}({\theta})$ (orange) with $x=12~\mbox{mm}$. The dashed lines show the beam envelope due to the feedhorn far-field beam pattern.}
   \label{figidealbeam}
\end{figure}

The response measured at a given point $\textbf{r}$ in the focal plane, as a point source in the far-field is scanned, we call the synthesized beam $B_{\textbf{r}}(\textbf{n})$. Since in QUBIC the sky-facing and detector-facing horns are identical, for an ideal optical combiner the PSF and the synthesized beam are equivalent, i.e. $B_{\textbf{r}}(\textbf{n})=\mbox{PSF}_{\textbf{n}}(\textbf{r})$ if $n_x=x/f$ etc., assuming $f$ is large enough to use the small-angle approximation. Figure~\ref{figidealbeam} shows the ideal synthesized beam for a point at the centre of the focal plane and for another at $x=12~\mbox{mm}$ from it. Moving off-axis causes the peaks to shift under the feedhorn beam envelope and the intensity of the pattern is further reduced by the beam intensity at an angle of $x/f$.  The width and position of the peaks in the synthesized beam scale with wavelength and this can be exploited in a technique called spectral-imaging~\cite{2020.QUBIC.PAPER2}.

\subsection{Optical design}{\label{subsec:combinerdesign}}

QUBIC was designed to observe at 150~GHz, in the first instance, and sample the range of multipoles $l\approx30-200$ in the sky. A 220-GHz band was added later to improve astrophysical foreground subtraction. The practical limitation on the size and number of bolometers that could be produced for the focal plane, as well as the requirement to Nyquist sample the fringes from the largest multipoles, meant that the focal length of the combiner was restricted to approximately 300~mm or less.
Since the input aperture had to accommodate an array of 400 feedhorns, it too is of this order resulting in a very fast optical system. We aimed to minimize the size of the cold components and, for logistical reasons, the full combiner had to fit into a cryostat with a volume of approximately $1~\mbox{m}^3$.

\begin{figure}
\centering
\includegraphics[ width = 0.8 \hsize ]{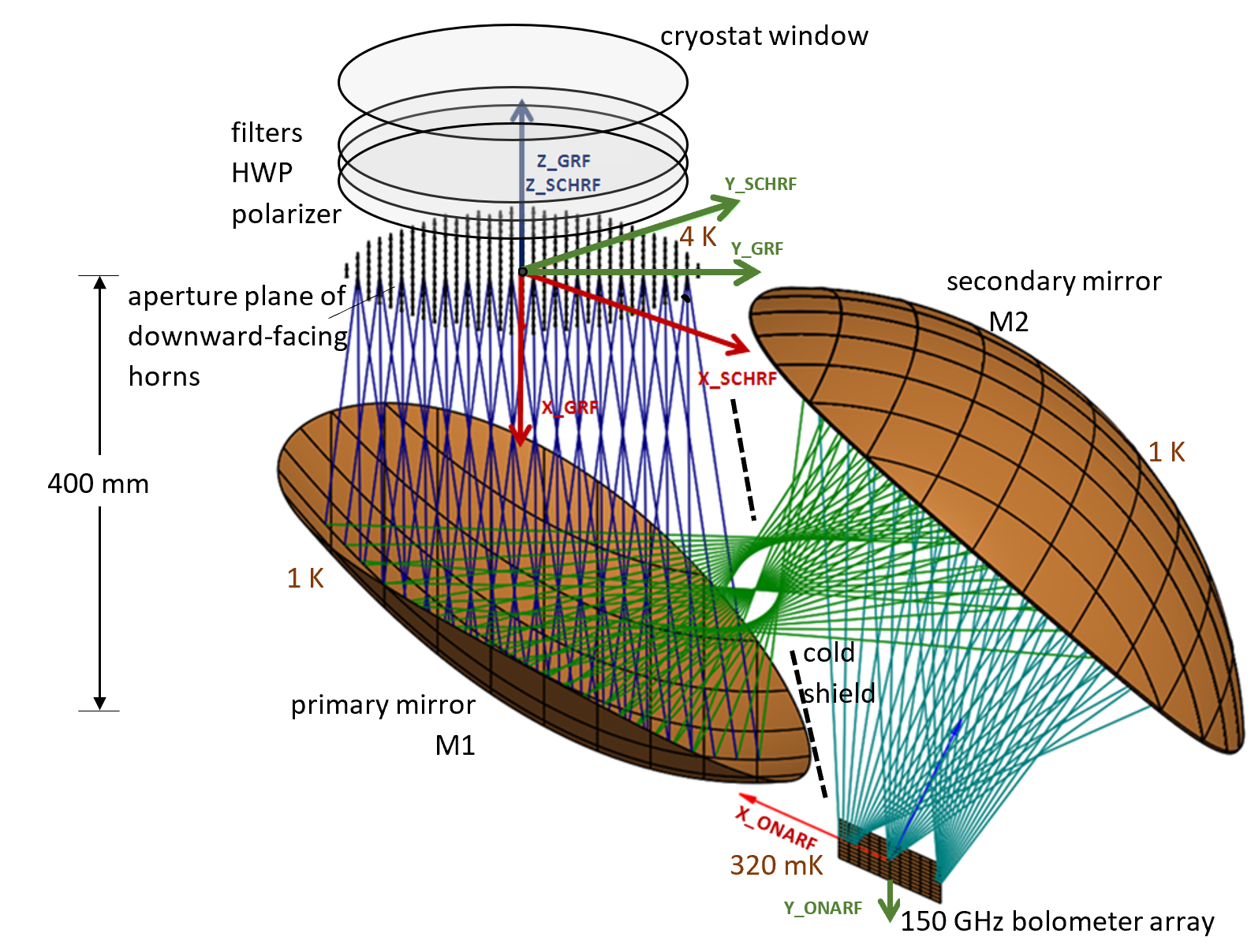}
   \caption{Schematic diagram of the QUBIC FI optical combiner. Rays were generated using the \swname{GRASP} software package. In the figure we omit the dichroic that will ultimately be located in front of the FI 150-GHz focal plane to reflect the 220-GHz radiation onto an orthogonal TES array. Some of the coordinate reference frames used in setting up the optical model are shown e.g. the global reference frame (GRF) and the on-axis focal plane reference frame (ONARF).} 
   \label{figcombinerraytrace}
\end{figure}

\begin{figure}
\centering
\includegraphics[ width = 0.6 \hsize ]{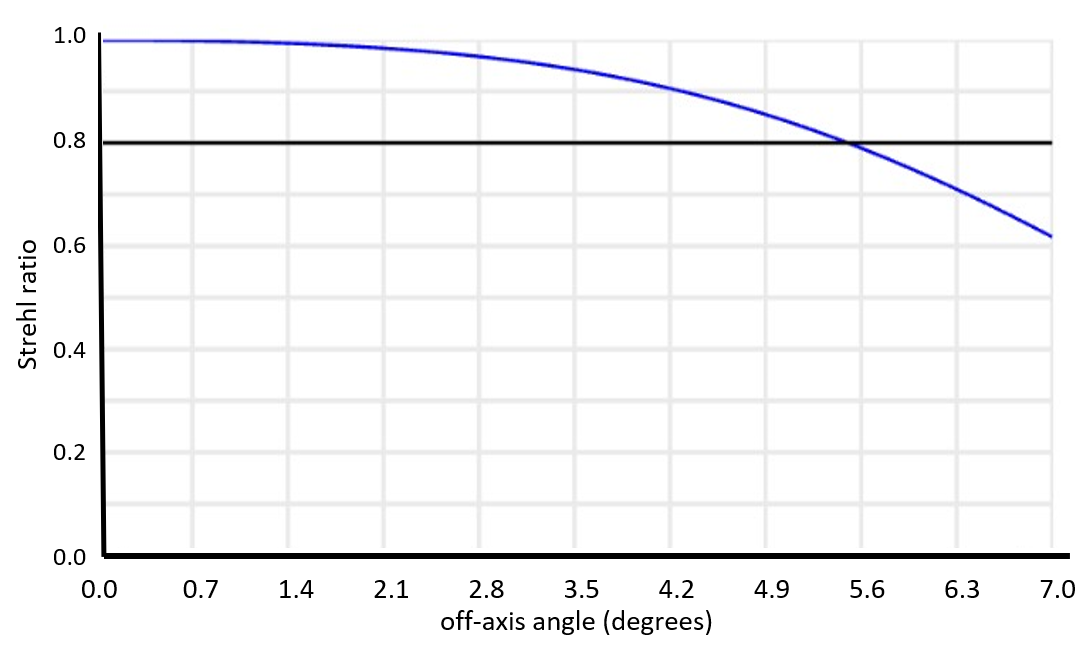}
   \caption{The strehl ratio achieved for the optical beam combiner (field-of-view  $\pm6.5^{\circ}$). This plot was produced using \swname{Zemax} software.} 
   \label{figstrehlratio}
\end{figure}

We preferred a reflective over a refractive system because its behaviour is well-understood and can be accurately modelled with PO software. We chose an off-axis design because of the large unobstructed aperture that this concept can offer.  This may change for any future modules as lens manufacturing and modelling improve.  Space limitations restricted us to dual-mirror options. We studied several designs for the beam combiner including compensated classical Cassegrain, Gregorian and Dragonian dual reflectors \cite{BennettPhD}. Both standard and crossed (front- and side-fed) \cite{Tran2008} geometries were considered but the low f-number ruled these out. Finally, a compensated off-axis Gregorian design (figure~\ref{figcombinerraytrace}) was chosen that obeyed the Mizuguchi-Dragone condition for minimum cross-polarization and astigmatism \cite{1978.Mizuguchi},\cite{1978.Dragone},\cite{1990.Rusch}. A further optimization of the mirror surfaces was carried out with the aid of commercial ray-tracing software (\swname{Zemax}\footnote{\url{https://www.zemax.com/}}, figure~\ref{figstrehlratio}) to improve the diffraction-limited field-of-view. The design is close to telecentric (within approximately -3$^{\circ}$ to +10$^{\circ}$ over the focal plane).

The cold HWP and polarizer are placed in front of the sky-facing horns. A dichroic filter, placed between the secondary mirror and the focal plane, transmits the 150-GHz band and reflects the 220-GHz band onto a separate focal plane. Each focal plane is tiled with 1024 NbSi TESs (32 of which are “dark” pixels that are outside the focal plane and are used for calibration), in four quadrants, cooled to 320~mK and read out with time-domain multiplexing \cite{2020.QUBIC.PAPER4}.

\subsubsection{Component sizes}{\label{subsubsec:modsize}}

\begin{figure}
   \centering
   \includegraphics[ width = 0.7 \hsize ]{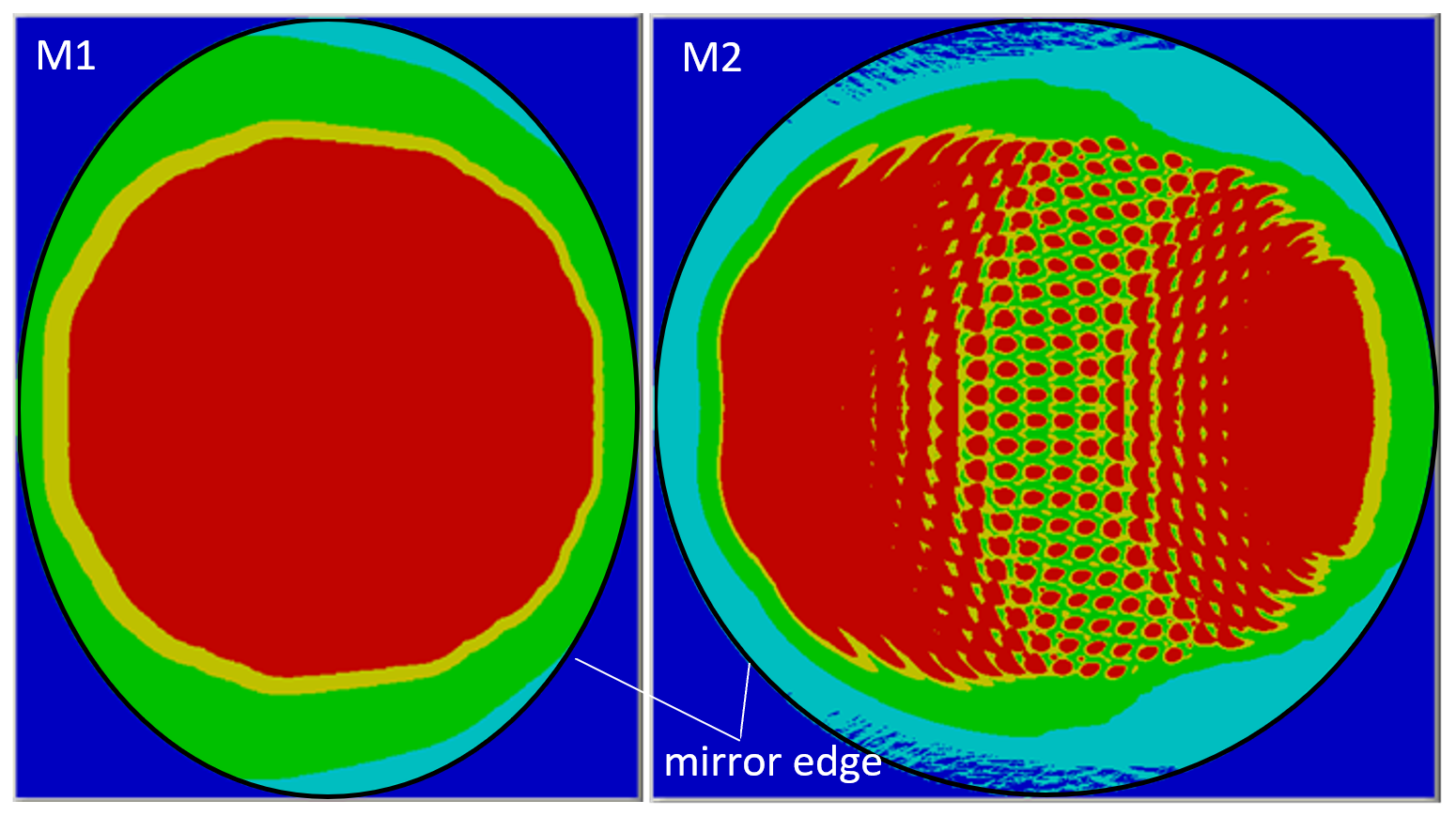}
      \caption{Beam footprints on (left) the FI primary mirror, M1, and (right) the FI secondary mirror, M2, \cite{ScullyPhD}. Each beam is coloured so that red corresponds to the region where 72\% of its power lies (the power expected to reach the focal plane), and yellow, green and pale blue to 86\%, 99\% and 100\%, respectively. On the primary mirror the beams are wide and overlap; on the secondary mirror the beams are generally narrower (see figure~\ref{figidealrealray}) and their location reflects the layout of horns in the aperture array.
              }
         \label{footprint}
   \end{figure}

\begin{figure}
   \centering
  \includegraphics[width= 0.5 \hsize]{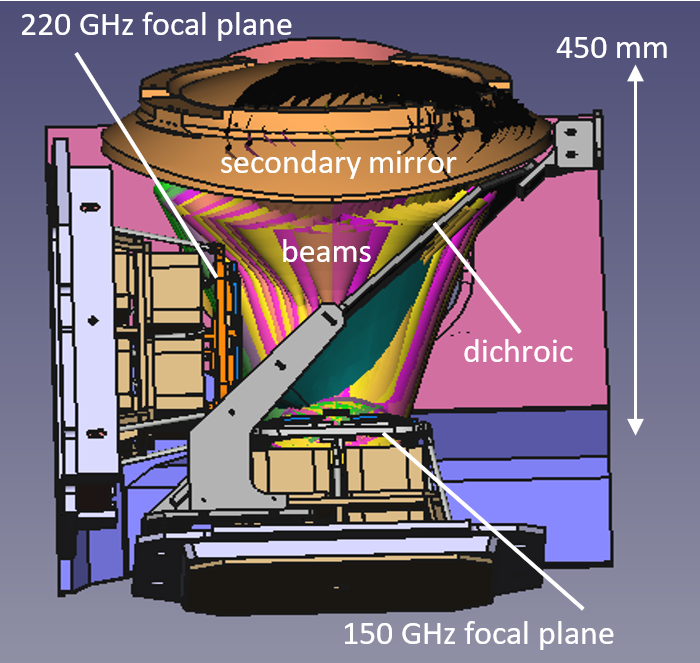}
      \caption{CAD model of the beam combiner showing the 3D model of the beams (here coloured yellow pink and green) propagating between the two mirrors \cite{GayerSPIE2016},\cite{ESAAntennas}.
              }
         \label{3DBeams}
   \end{figure}

Minimum component and aperture sizes are calculated by PO propagation of beams to different planes in the instrument using either the industry-standard software \swname{GRASP}\footnote{\url{https://www.ticra.com/software/grasp/}} or our in-house software \swname{MODAL} \cite{GradzielSPIE2007}. We then use footprint diagrams ($e.g.$ figure~\ref{footprint}) at the component of interest or reconstruct three-dimensional beams from their edges calculated at a series of planes throughout the instrument (figure~\ref{3DBeams}, and see also Gayer \textit{et al.} \cite{GayerSPIE2016}).  The edge is defined as either the points at which the beam intensity has dropped to a certain level or, for the more complex beam shapes close to focus or for the multimode beams in the higher band, we use encircled energy. The beam edges can be imported into a CAD model of the system to aid in its design. The wider beams in the lower band determined the minimum component sizes.

\subsubsection{The 1K box}{\label{subsubsec:1kbox}}
The optical combiner elements and the detectors are placed inside the cryostat \cite{2020.QUBIC.PAPER5} on a support structure dubbed the  “1K box” (due to the temperature to which it will be cooled). The 1K box is a complex aluminium structure built to house and hold in place all the components in the optical chain with the minimum structural mass possible for ease of cooldown. To avoid distortions in the optical path caused by the linear shrinking of the mirror surfaces during cooldown, neither mirror is rigidly anchored to the 1K box but is held in place by aluminium hexapods of varying and adjustable length mounted on support rings, see figure~\ref{fig_hexapods} (left) in section~\ref{subsubsec:aliCombiner} .

\subsubsection{Stray-light}
\begin{figure}
   \centering
  \includegraphics[width=0.7\hsize]{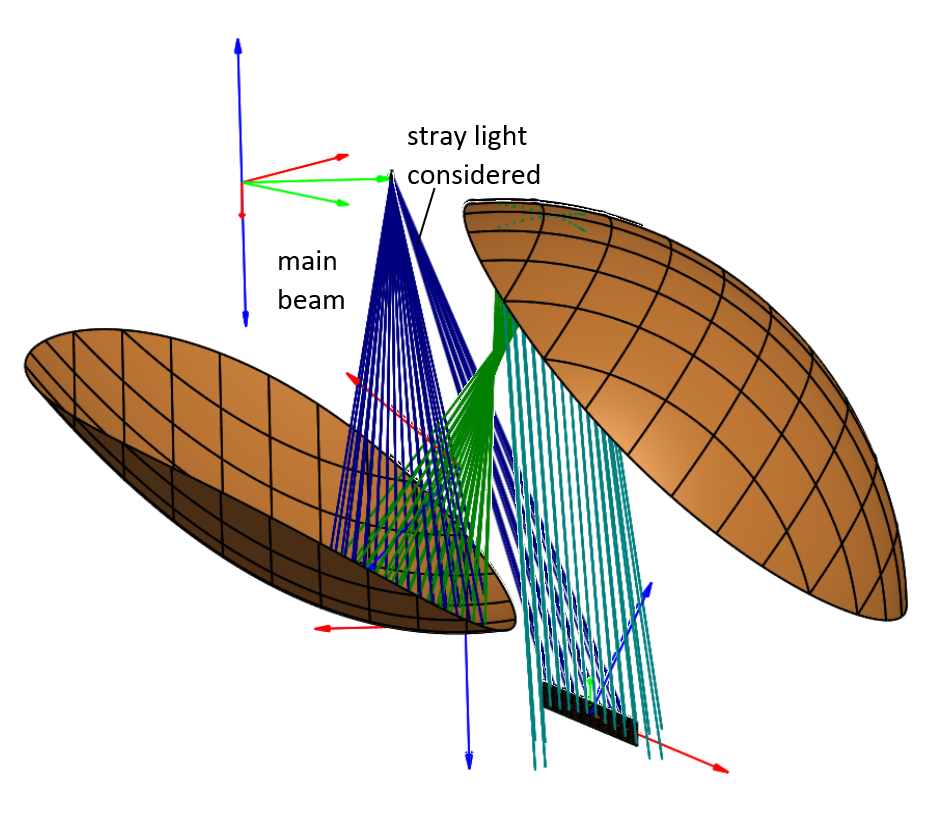}
      \caption{The optical paths considered in the stray-light analysis. Rays propagating through the FI combiner (as intended, green) and the ones directly illuminating one focal plane are shown. The stray light analysis was carried out for the rays directly illuminating the focal plane (narrow, blue ray bundle). In this worst case scenario, no cold stop and shielding around the dichroic are present.
              }
         \label{fig_straylight}
   \end{figure}

{\label{subsubsec:modstray}}   
Radiation from the main beam of the detector-facing feedhorns is reflected from the combiner mirrors before reaching the focal plane. At large angles in some specific directions, however, stray-light could reach the focal plane directly, as shown in figure~\ref{fig_straylight}. For this study neither the cold stop nor the dichroic were included, so the worst case scenario has been considered. We used \swname{GRASP} PO to estimate this stray-light contribution from the feedhorn array, in terms of contaminant, that could be added to sky fringe patterns and the resulting increased radiative power loading. Pairs of feedhorns were chosen as test cases, selecting horns at the edge of the array most likely to cause a problem. We found that the amplitude of fringes generated by radiation that reached the focal plane directly was at least four orders of magnitude lower than that from radiation propagating through the combiner and at different spatial frequencies.  Even in the case of the edge feedhorns, the power collected by the 150-GHz focal plane from direct radiation is almost three orders of magnitude lower than that from the combiner. Inclusion of the cold stop and the shield around the dichroic filter will further reduce these effects. 
   
 A final stray-light analysis was carried out using ray tracing and \swname{Zemax}'s non-sequential mode to ensure there were no optical paths of particular concern.

\subsection{Synthesized beam and PSF modelling}{\label{subsec:modelling}}

The initial design and final optimisation of the dual-reflector design was carried out using ray-tracing in order to take advantage of the speed and optimisation routines available in the commercial software package \swname{Zemax}. However, for a detailed analysis at the operating frequencies, where component sizes are not very large compared with the wavelength of radiation, techniques that include the effects of diffraction were used. Our most accurate models use PO to propagate beams through the combiner (e.g. using the software package \swname{GRASP}) and electromagnetic mode-matching to model coupling to, and the beam patterns from, the back-to-back feedhorn array. When compared with the analysis of an ideal system in section~\ref{subsec:combinerideal}, this PO analysis includes the exact shape of the horn beam pattern as well as any truncation, aberration or cross-polarisation introduced by the real components of the combiner. The short focal length in particular is expected to introduce aberrations into the instrument PSF. 

\begin{figure}
   \centering
  \includegraphics[width= 1.0 \hsize]{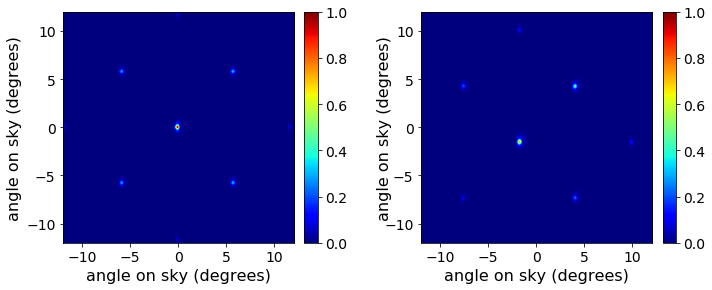}
  \includegraphics[width= 1.0 \hsize]{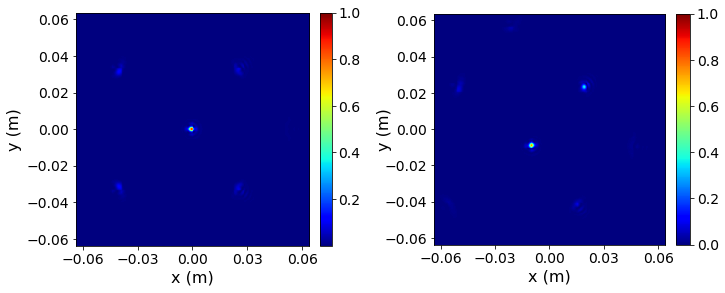}
      \caption{(top) The simulated 150-GHz synthesized beams of QUBIC (top left) for an on-axis point source on the focal plane and (top right) for a point source that is 12~mm off-axis. The central peak moves 2.29$^{\circ}$ from the centre of the sky for this off-axis focal-plane point and the relative amplitude of the peaks change. (bottom) The simulated PSF of QUBIC (bottom left) for an on-axis point source in the sky and (bottom right) for a point source that is 2.29$^{\circ}$ off-axis at 150~GHz. The PSF is the focal plane image. The central peak moves 12~mm from the centre of the focal plane for this off-axis source and the relative heights of the peak change. These simulations were carried out using vector PO and electromagnetic mode matching for the horns; they include the effects of aberrations and truncation by the combiner. The horn array was rotated by 45$^{\circ}$ with respect to the focal plane coordinate system ONARF (shown in figure~\ref{figcombinerraytrace}).  
              }
         \label{SBsandPSFs}
   \end{figure}

\begin{figure}
   \centering
   \includegraphics[width=0.49\linewidth]{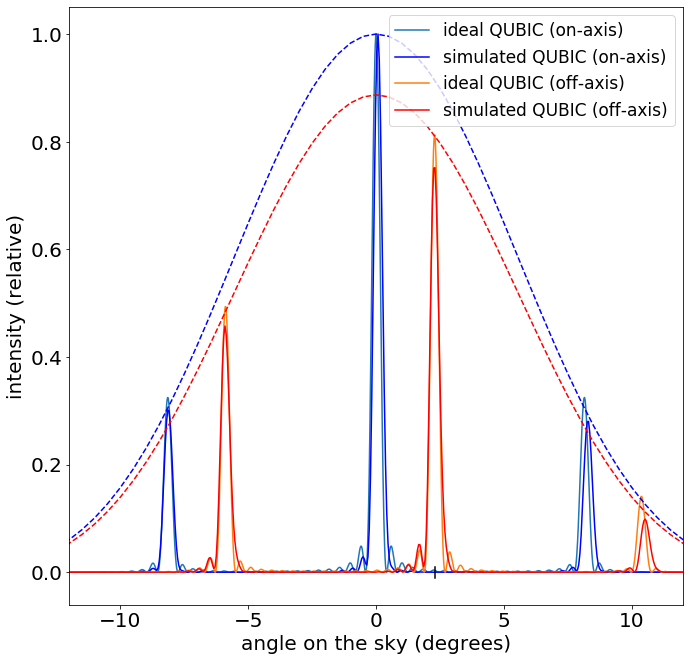}
   \includegraphics[width=0.49\linewidth]{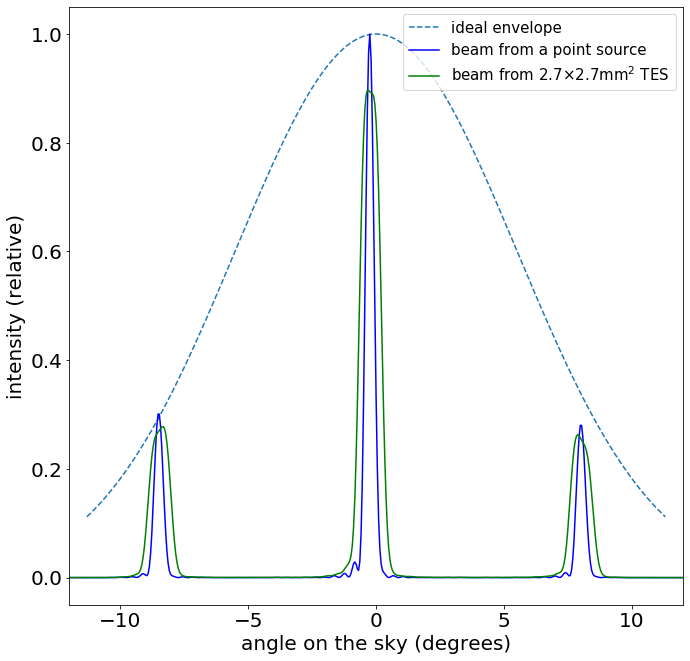}
      \caption{(left) Comparison of the simulated synthesized beam at 150~GHz for a realistic and ideal QUBIC beam combiner for the two focal plane positions of figure~\ref{figidealbeam}: $x=0$ and $12~\mbox{mm}$. The envelope of the simulated horn beam is shown, rather than the best-fit Gaussian beam. In the far-field, the main effect of the aberrations is on the height and position of some peaks rather than on their width. (right) Simulated synthesized beam at 150~GHz for an on-axis point source on the focal plane and for a hypothetical on-axis TES of finite area. (In QUBIC none of the TESs is exactly on-axis.) The beam envelope for an ideal combiner is shown for comparison. }

         \label{SBEAMs&TESbeams}
   \end{figure}

We simulate the synthesized beam of a particular TES by using PO to propagate the beam from a point source at its location on the focal plane to the secondary mirror, through the cold shield aperture (which we call the cold stop), onto the primary mirror and then to the feedhorn array aperture plane.  We then use electromagnetic mode matching to determine the set of TE and TM modes excited in each feedhorn by this beam and to propagate them through the corrugated horns. From the feedhorn aperture we again use PO to propagate the beams through the filter and cryostat window apertures, and on to the sky. The coherent addition of the beams from each of the feedhorns gives the synthesized beam of the TES. Figure~\ref{SBsandPSFs} (top) shows a plot of the synthesized beam at 150~GHz for an on-axis point source on the focal plane as well as for a point source displaced 12~mm from it. Cuts through the peaks are compared with the prediction for an ideal combiner in figure~\ref{SBEAMs&TESbeams} (left). The aberrations introduced by the realistic combiner (i.e. the off-axis model including aberrations and truncation) cause changes in the height of some peaks, especially for off-axis TESs but there is good agreement with the predicted peak positions and widths. At 150~GHz the FWHM of the peaks is 0.4$^{\circ}$ and the secondary peaks are 8.2$^{\circ}$ from the main peak. The off-axis nature of the combiner causes an asymmetry in the small features either side of the peaks. The envelope of the synthesized pattern, and therefore the relative height of the peaks, is determined by the far-field beam pattern of the individual horns (figures~\ref{figbeam150} and \ref{figbeam220}). The finite integration area of the TES is simulated by repeating the procedure above from a grid of nine points in the TES area on the focal plane ($2.7~\mbox{mm} \times 2.7~\mbox{mm}$) and adding the resulting intensities (figure~\ref{SBEAMs&TESbeams} (right)). This analysis assumes that each point on the TES absorber acts independently and with equal sensitivity to the incoming field.

To calculate the instrument PSF we start by coupling a plane wave, truncated by the cryostat window and filter apertures, to the sky-facing feedhorn apertures, use electromagnetic mode-matching to determine the beams emitted by the detector-facing feedhorns and then use PO to propagate them through the optical system (in the order: primary mirror, cold stop, secondary mirror) and on to the focal plane. Truncation by the apertures has negligible effect at 150~GHz, where the horns are single moded, so we can omit the first step and begin the simulation from the previously calculated detector-facing feedhorn beams with the appropriate phase slope if the source is off-axis. At 220~GHz, where the horns are multi-moded, coupling to the modes for different angles of incidence varies in a complex way and so we carry out the full simulation with \swname{MODAL} which combines electromagnetic mode-matching with PO. Aberrations in the combiner produce some distortion in the peak locations on the focal plane (e.g. the asymmetry in the secondary peak location with respect to the main peak in figure~\ref{SBsandPSFs} (bottom)). The short focal length of the combiner makes the aberrations in the PSF appear more pronounced than in the sky (far field). The PSF is used for comparison with laboratory measurements during instrument calibration campaigns \cite{2020.QUBIC.PAPER3}. The same simulations show that the initial 0.06\% cross-polar component of the 150-GHz feedhorn beam increases to about 0.5--1.9\% after propagating through the combiner, depending on where the horn is in the aperture array.  This cross-polarization, since it is introduced after the HWP and polarizer, does not affect the separation of Q and U signals.

\subsubsection{Performance requirements}{\label{subsubsec:modwindow}}

The performance of the QUBIC instrument design is quantified using the window function (WF), a measure of its sensitivity to different multipoles on the sky.  If the synthesized beam pattern for a position on the focal plane, $\textbf{r}$, is written in terms of spherical harmonics as

\[
B_{\textbf{r}}(\textbf{n})  =\sum\limits_{lm} \beta_{lm,\textbf{r}}(\textbf{n})Y_{lm}(\textbf{n})
\]

\noindent
where $\textbf{n}$ is the direction in the sky relative to the telescope pointing centre, $l$ and $m$ are the usual degree and order number, respectively, and if the sky image in intensity is decomposed into the same spherical harmonics as 

\[
I(\textbf{n})  =\sum\limits_{lm} a_{lm}Y^{*}_{lm}(\textbf{n})
\]

\noindent
then the image in total intensity $I$, is

\[
S_{I,\textbf{r}}(\textbf{n})  =\sum\limits_{lm} a_{lm}\beta_{lm,\textbf{r}}(\textbf{n})
\]

\noindent
where $a_{lm}$ are the spherical harmonic coefficients and, for the Gaussian fluctuations of the CMB,  $\langle a_{lm}a_{l'm'} \rangle =C_{l} \delta_{ll'} \delta_{mm'} $.  We then define the window function as

\[
W_{l,\textbf{r},\textbf{r}'}(\textbf{n}\textbf{n}')  =\sum\limits_{m} \beta_{lm,\textbf{r}}(\textbf{n}) \beta^*_{lm,\textbf{r}'}( \textbf{n}' ).
\]

\noindent
Similar window functions can be defined for images in the other Stokes’ parameters $Q$ and $U$. Because it is an interferometer, window functions for QUBIC show multiple peaks and examples have been shown in Battistelli \textit{et al.} \cite{Battistelli2011}. The simulated window function over the multipole range $\ell=30-200$, relative to that of an ideal instrument without aberration or truncation, is used as a figure-of-merit for our combiner designs and in our tolerance analyses.

\begin{figure}
   \centering
   \includegraphics[width = 0.6 \hsize]{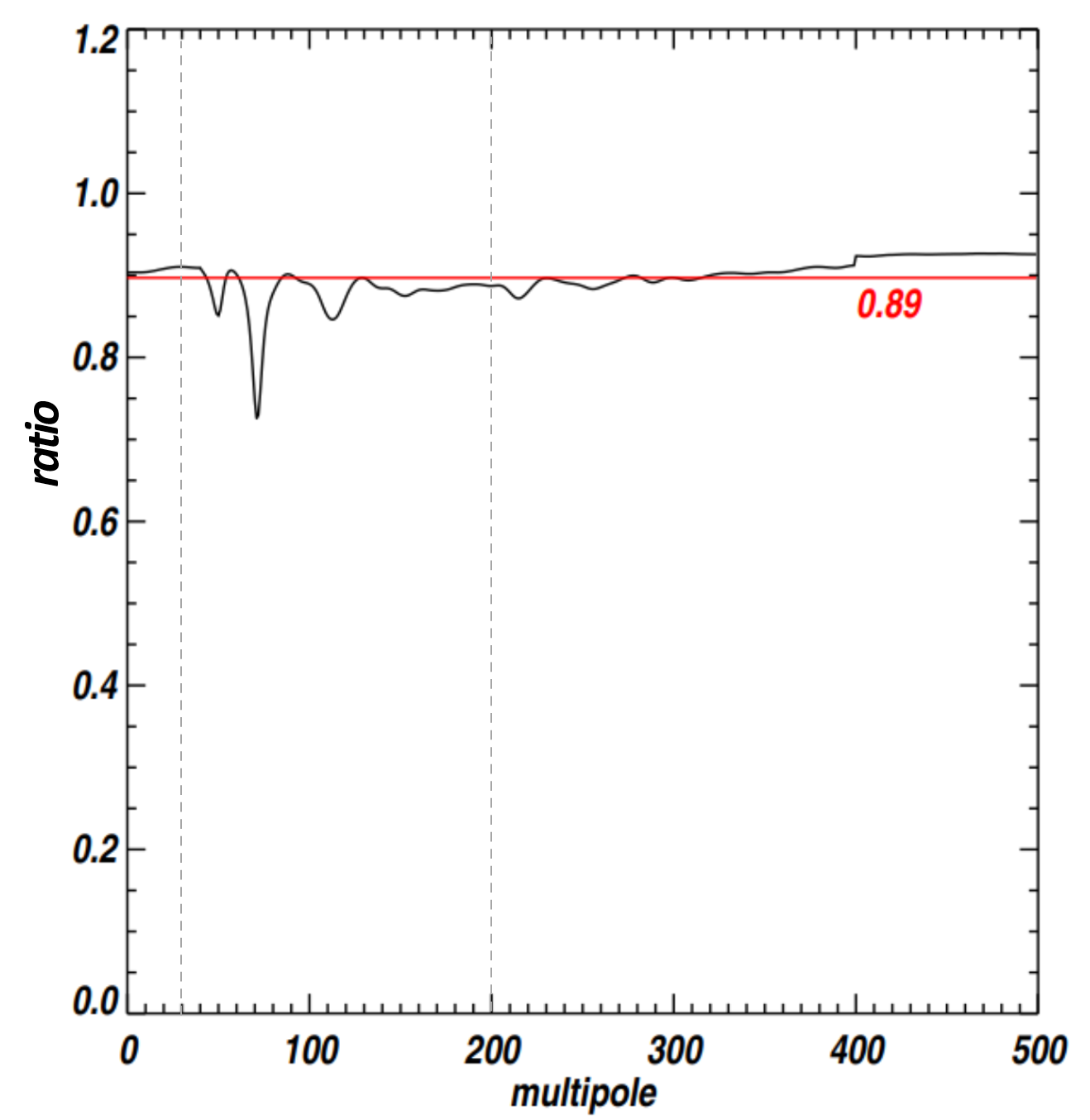}
      \caption{Diagonal window function ($W_{l,\textbf{r},\textbf{r}}(\textbf{n},\textbf{n}) $) of the simulated real beam combiner divided by that of an the ideal non-aberrating instrument (black) \cite{BigotSazyPhD}. The average ratio of 0.89 is shown by the red line. Features in the ratio arise due to the interferometric nature of QUBIC \cite{Battistelli2011}. The multipole range of interest here, $\ell=30-200$, is indicated by the dashed lines. 
              }
         \label{fig_windowfunction}
   \end{figure}

By using PO simulations to find the synthesized beam, we have calculated the WF sensitivity, WFS, that is the diagonal window function of our final beam combiner design and compared it to that of an ideal one. The result, see figure~\ref{fig_windowfunction}, shows that the effect of the aberrations and truncations is to reduce the sensitivity of the instrument by $\sim10\%$. In section~\ref{subsubsec:beam measurements} we compare measured and simulated synthesized beams directly.

\subsection{The technological demonstrator}{\label{subsec:TD}}
The QUBIC TD was built to test and validate the QUBIC design and all its sub-systems (optics, cryogenics, detectors, read-out electronics, scan strategy and calibration). It differs from the FI in that it has a reduced number of detector pixels (256 pixels, 248 of which are inside the focal plane, i.e. one quarter of one focal plane), a reduced number of feedhorns (the central 8×8 horns of the full array, figure~\ref{hornlayout}), reduced primary and secondary mirror size (400~mm in diameter rather than 600~mm), reduced HWP (180~mm) and filter sizes (up to 280~mm in diameter) and a neutral density filter, instead of a dichroic, to manage the radiative loading inside the laboratory on the single focal plane. In all other respects the TD is the same as the FI.

\subsubsection{Manufacturing of the mirrors}{\label{subsubsec:manuCombiner}}

The mirror surfaces that were designed as presented in section~\ref{subsec:combinerdesign} were the surfaces required at the cryogenic operating temperature of 1~K. In order to manufacture the mirrors at 300~K, the designs were modified using the software CATIA\footnote{\url{https://www.3ds.com/it/}} assuming a linear thermal expansion, integrated over the range 1~K -- 300~K, of $\Delta L/L=-0.004348$, to give what we name the “nominal design” at 300~K.  

The aluminium alloy 6082-T6 was selected for the mirrors because of its excellent thermal and mechanical performance. For the purpose of both modelling and manufacturing, the surfaces are described by a quadric:
\begin{equation} \label{eq:quadric}
  Ax^2+By^2+Cz^2+Dxy+Exz+Fyz+Gx+Hy+Iz+J=0
\end{equation}
with parameters appropriate for both mirrors and for two temperatures: 1~K (design and operation temperature) and 300~K (manufacturing temperature). To model the nominal mirror designs at both temperatures, we imported them into the \swname{Zemax} and \swname{GRASP} software packages as STP-type files with clouds of points representing their surfaces. In this way we could optically model the ideal as-designed instrument. The same STP files were used as the blueprints for the milling machine control software.

\begin{figure}
   \centering
   \includegraphics[width = 0.9 \hsize]{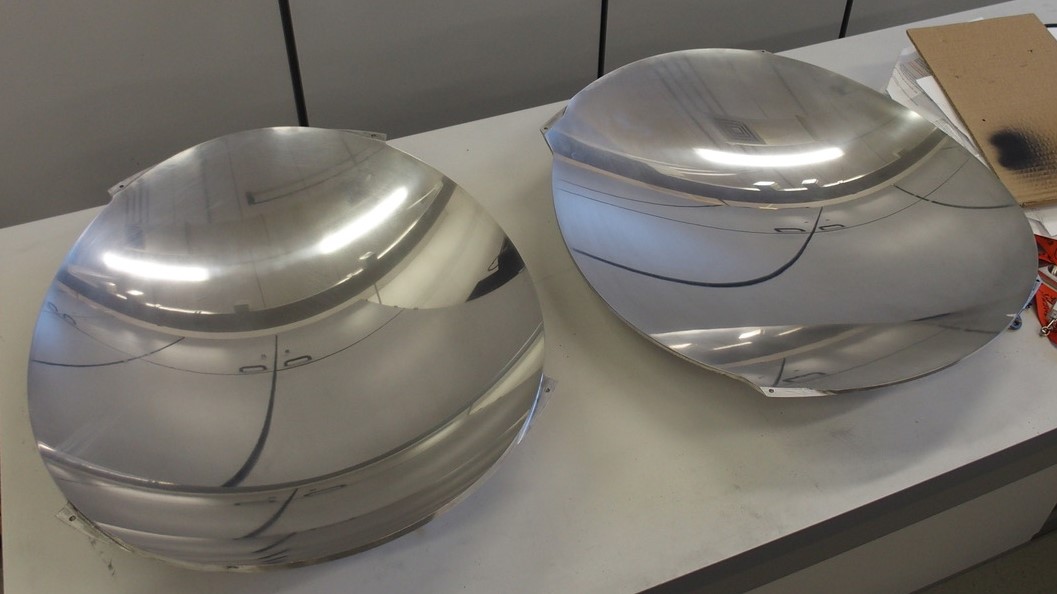}
      \caption{The two mirrors of the FI combiner during the manufacturing procedure, before the final polishing step.
              }
         \label{fig_FI_mirrors}
   \end{figure}

The mirrors for the TD were manufactured in the Milano-Bicocca University machine workshop using a three-axis GB Ferrari milling machine. The manufacturing process included cycles of thermal treatments between 70~K and 610~K to relieve internal stress. The FI mirrors were manufactured by 3C S.r.l.\footnote{\url{http://www.3c-stampi.it/}} using a FIDIA gantry-type high-speed milling machine. In figure~\ref{fig_FI_mirrors} the two mirrors of the FI are shown before the last polishing procedure.

\subsubsection{Mirror metrology}{\label{subsubsec:measuMirror}}

3D metrology on both TD mirrors was performed in a temperature-controlled room at the Sezione INFN machine workshop in Rome using a Poli Galaxy Diamond measuring machine with a ruby probe. Measurements were repeated over several days, testing both the mechanical stresses imposed by the rigid mounting structure and the capability of the machine to correctly map the surfaces in the presence of arbitrary tilts and displacements of the entire mirror surface. The data produced by this 3D metrology (ANSI and STP file types) were then used to check the manufactured mirror profiles against the 300 K nominal design that were also described by a cloud of points in an STP file. \swname{ANSYS R14.5}\footnote{\url{https://www.ansys.com/}} software was used to model the contraction of the measured surfaces down to the cryogenic temperature of 1 K.
This cloud of cryogenic surface points were incorporated into the \swname{GRASP} PO model of the TD optics and the performance of the manufactured combiner was tested against the ideal as-designed surfaces using the calculated WF  (section~\ref{subsubsec:modwindow}). The ratio of the real to the nominal WF was $95\%$, satisfying the estimated requirements.

The FI mirrors were measured by Media Lario\footnote{\url{https://www.medialario.com/}} with a Poli TCX coordinate measuring machine and followed the same pipeline of consistency checks that were developed for the TD.

\subsubsection{Aligment of the TD combiner}{\label{subsubsec:aliCombiner}}

\begin{figure}
   \centering
   \includegraphics[width = 1 \hsize]{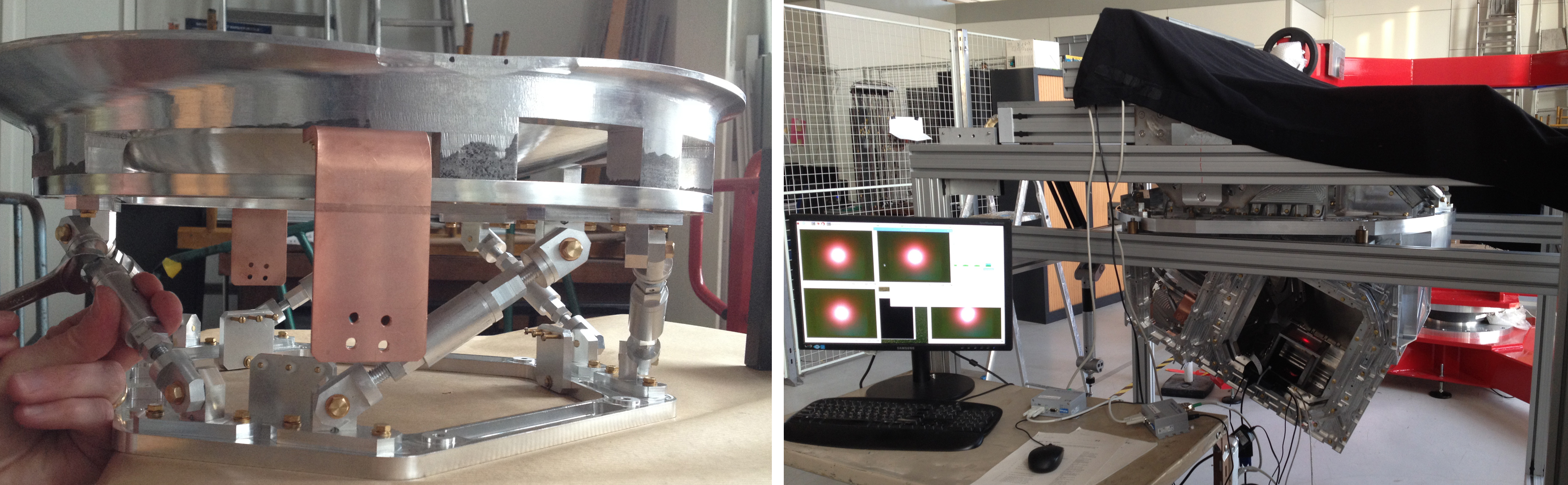}
      \caption{(left) TD primary mirror with the hexapods supporting and moving it on the backside and (right) the combiner inserted in the "cart" during He-Ne laser alignment at different elevation angles at APC.
              }
         \label{fig_hexapods}
   \end{figure}

Each of the mirrors in the combiner was mounted on its supporting hexapod (see section~\ref{subsubsec:1kbox})  and its position verified using FARO arm 3D measurements (Bras Measure 3D Romer). Figure~\ref{fig_hexapods} (left) shows the TD primary mirror supported by hexapods.

To ensure the correct positioning of the mirror surfaces, we planned a procedure to check the alignment of the combiner at ambient temperature by illuminating it with an optical He-Ne laser, essentially reproducing a ray-tracing analysis. Automated movements allowed the laser to scan a grid pattern reproducing the 8 $\times$ 8 feedhorn array positions. At each of the 64 steps a CCD camera recorded an image of the light distribution on a screen placed at the nominal position of the focal plane. For comparison, we generated optical models of the combiner at ambient temperature with \swname{Zemax}, using the real measured surfaces of both TD mirrors (see section~\ref{subsubsec:measuMirror}). 

The alignment tolerance specifications were set by a tight requirement of no more than $1\%$ decrease in the WFS (see section~\ref{subsubsec:modwindow}). This corresponds to misalignments of $\pm1$~mm or rotations of $\pm0.1^{\circ}$ up to $\pm0.5^{\circ}$ around the normal axis at the centre of the mirror section \cite{OSullivanSPIE2018Optics}. The laser mount was moved using two VT-80 single-axis 300-mm long translation stages from PI miCos. The images on the focal plane screen were taken with a Raspberry PI V2 camera and analysed with custom IDL scripts. The alignment procedure involved both “on-axis” scans (with the laser pointing straight, parallel to the feedhorn optical axes) and “tilted” scans (with the laser tilted by $\pm6.5^{\circ}$ with respect to two orthogonal directions to reproduce the distribution of the light within the FWHM of each feedhorn beam). Furthermore, the combiner and the laser were mounted on a supporting structure, the “cart”, that allowed us to tilt the whole optical ensemble in order to study any change in alignment during the planned elevation scans. Figure~\ref{fig_hexapods} (right) shows the combiner during the test at APC in Paris. 

The alignment was carried out in two different stages with several sets of measurements taken and analysed using three different pipelines to ensure the robustness of the results. The first two pipelines (“Roma” and “APC”) used a “centre-of-mass” style algorithm to compute the position of the centroid of the light produced in each individual image for a full scan in a given direction and then computed the mean position. The third pipeline (“Product”) assumed the light distribution in each image to be representative of the probability distribution function of the light on the focal plane, thus the product of the light distribution of all the images gave us the probability distribution of the mean centroid. The results of the procedure showed that we were able to recover consistent centroid positions with all our analyses and assuming nominal mirrors. However, we were not able to satisfy the tightest constraints of the \swname{Zemax} model ($1\%$ decrease of the WFS) due to a combination of poor mirror surface quality at the laser wavelength (a surface roughness of around 1~$\mu$m, in the worst case) and difficulty in making mechanical adjustments to the mirror positions in the combiner structure. By comparing with PO simulations, we were nevertheless able to verify that >$90\%$ of the power passing through the $8\times8$ feedhorn array was ending up at the correct position on the focal plane. The alignment procedure of the TD with the laser also allowed us to conclude, through elevation scans, that the combiner tilted rigidly and that there were no significant mechanical stresses on the focal plane. We were able to recover consistent centroid positions for scans up to a tilt of 60$^{\circ}$ in elevation. From this we concluded that the FI mirrors should have a finer polish and be made with reference marks on their rims to aid in their correct alignment with the FARO arm measurements.

\subsubsection{TD beam measurements}
\begin{figure}
   \centering
   \includegraphics[width = 0.9 \hsize]{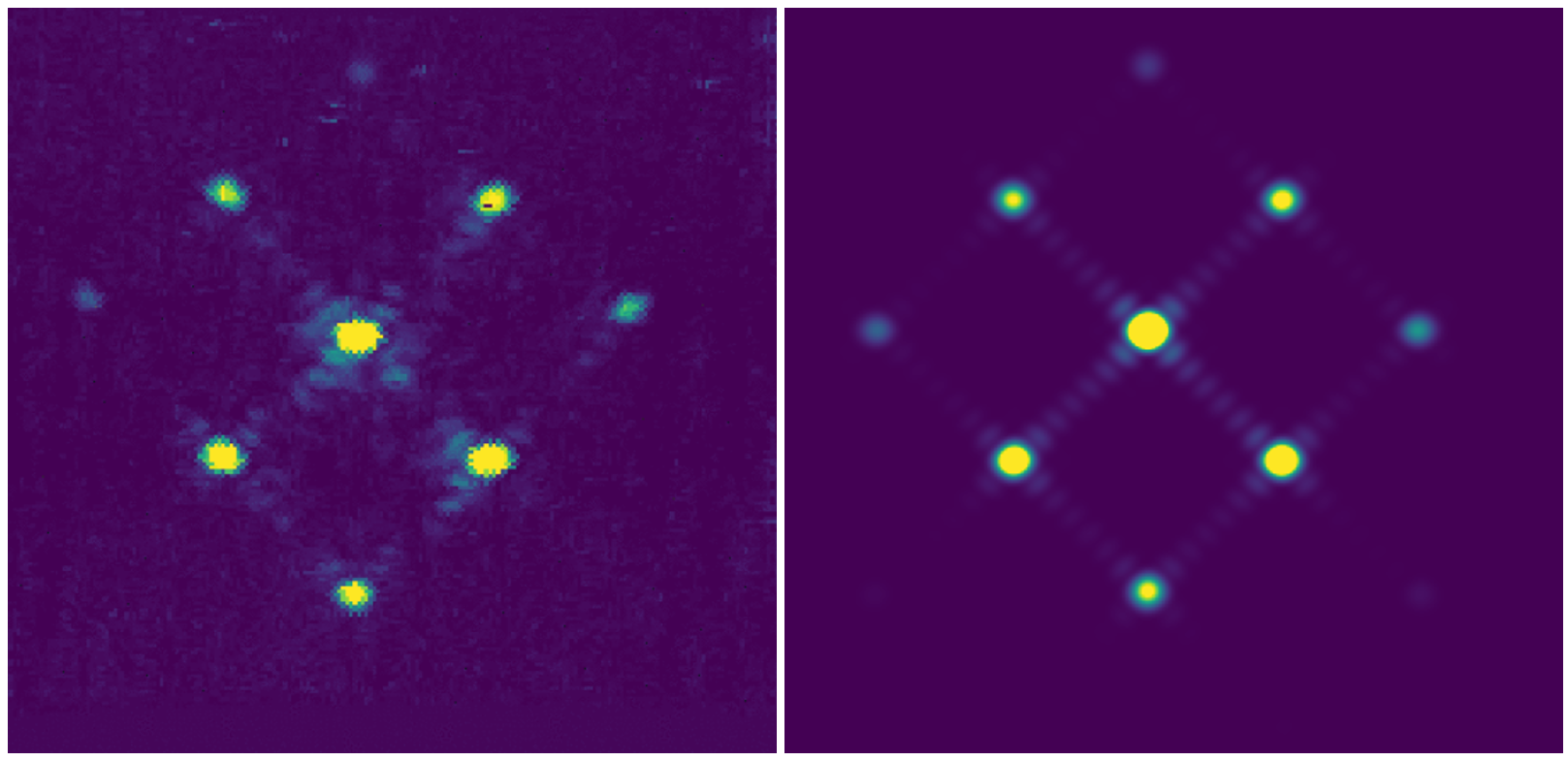}
      \caption{(left) Measured and (right) simulated 150-GHz TD synthesized beam pattern for a TES close to the focal plane centre. (This is the nearest equivalent of figure~\ref{SBsandPSFs} (top left) for the TD.) The TES signal for these measurements was saturated and so the colour range for the simulated pattern has been clipped for comparison. These plots show relative intensity on a logarithmic scale and cover an area of $33^{\circ} \times 33^{\circ}$\cite{EuCAP2020}.
              }
         \label{fig_measuredbeams}
   \end{figure}

{\label{subsubsec:beam measurements}}
Initial calibration of the TD was carried out at APC and is described in detail in Torchinsky \textit{et al.} \cite{2020.QUBIC.PAPER3}.  The aim of this first campaign was to measure the synthesized beam at multiple frequencies by scanning an external far-field calibrator source across the instrument field-of-view. The synthesized beam as measured by each TES was recorded. An example is given in figure~\ref{fig_measuredbeams} showing good agreement between the measured beam and the theoretical prediction, including aberrations.  The theoretical prediction was made by using full vector PO as described in section~\ref{subsec:modelling}.

\section{Groundshield and forebaffle}{\label{sec:GSFB}}

For high sensitivity ground-based millimetre observations the instrument must be shielded to reduce unwanted radiation from contaminants such as the Sun, Moon and ground, potentially reaching the detectors. A study of the shielding system for QUBIC was carried out in order to define its geometry and a suitable manufacturing material. The investigation was performed with \swname{GRASP} and \swname{CHAMP}\footnote{\url{https://www.ticra.com/software/champ3d/}}, combining MultiGTD (Geometric Theory of Diffraction) and MoM (Method of Moments) approaches to infer the pattern of the instrumental beam, including the impact of the shielding system, out to the far sidelobes at the lowest frequency band, centred at 150 GHz, where the requirements are estimated to be more demanding.

\begin{figure}
\centering
\includegraphics[width = 0.8 \hsize ]{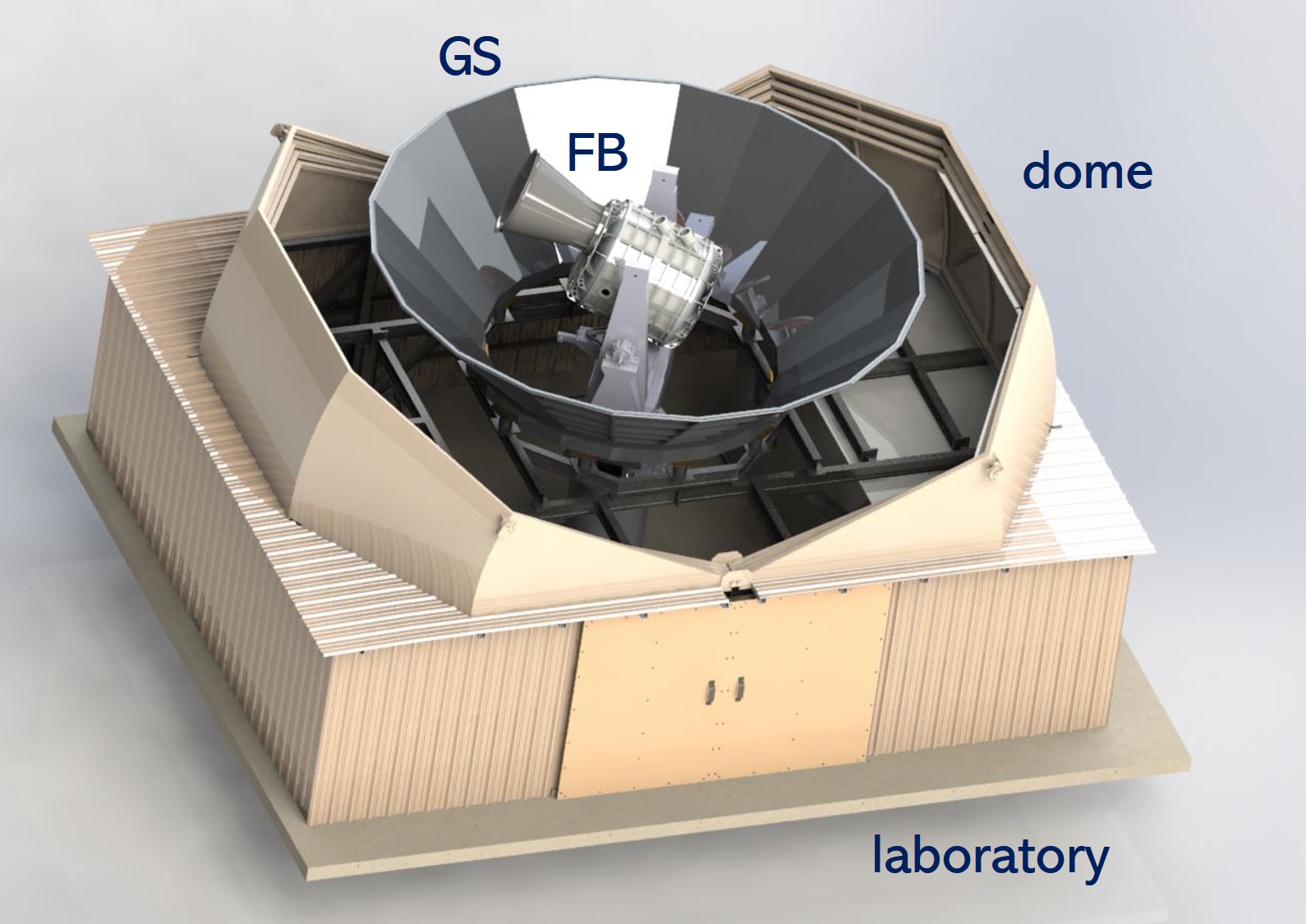}
\caption{Sketch of the QUBIC instrument, with the dome open, showing the cryostat with its FB on the altazimuth mount. These are enclosed inside the GS on the roof of the laboratory.}
\label{fig_lab_FB_GS}
\end{figure}

The shielding system has two components: the forebaffle (FB), a shield fixed to the cryostat window and so moving with the instrument optical axis, and the groundshield (GS), a shield fixed to the roof of the laboratory enclosing the instrument, see figure~\ref{fig_lab_FB_GS}. The insertion of a third intermediate shield (around the FB and also fixed to the cryostat window) was considered and studied but was then abandoned as it did not result in much improvement.

\subsection{Shield modelling}{\label{sec:modelFBGS}}

\begin{figure}
\centering
\includegraphics[width = 0.9 \hsize]{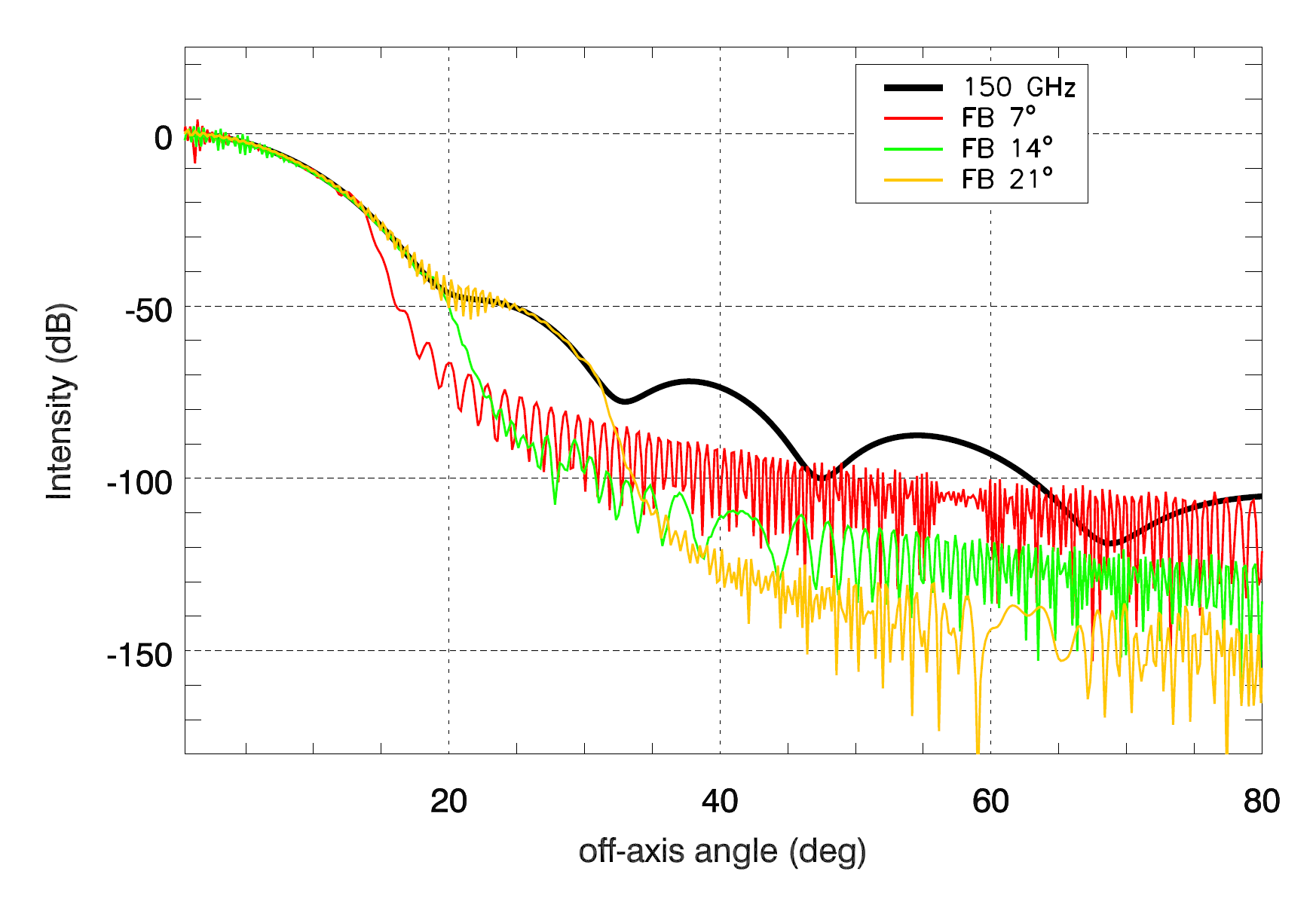}
\caption{Radial angular profile of the beam pattern for an on-axis feedhorn, modeled as a hybrid mode conical horn, without and with a reflective FB of 1 m height for three different semi-aperture angles. The colour code is specified in the legend.}
\label{figFB_aperture}
\end{figure}

In order to take a conservative approach, we have investigated the impact of the shielding configuration on the beam pattern of the central feedhorn of the array, at the frequency of 150~GHz, because it modulates the synthesized beam pattern.
Initially, we studied a cylindrical geometry for the FB but, mainly due to the dimensions dictated by the large angular aperture of the beam pattern (FWHW of $12.9^{\circ}$), we preferred to move towards a conical geometry. This geometry has been optimized by varying the height, from 50~cm up to 2~m, and the semi-aperture angle, from 7$^{\circ}$ (almost FWHM/2) up to 28$^{\circ}$ (almost 2FWHM), in order to minimize the sidelobes. Figure~\ref{figFB_aperture} shows the pattern comparison for three different angular apertures with the one resulting from the feed alone. In this case we used a simple hybrid mode conical horn while later we included the final feedhorn pattern as in figure~\ref{figbeam150}. We rejected solutions with a semi-aperture angle larger than 14$^{\circ}$ because they offered no advantages, even from a mechanical manufacturing point-of-view.

A fully reflective solution for the internal surface of the FB was compared with an absorptive one to highlight the main differences on the beam pattern and to choose the material of the inner surface. The MultiGTD approach does not allow us to analyze reflectors covered by dielectric materials with defined electrical properties. To overcome this limitation, we have performed our simulations with the help of \swname{CHAMP}, which allows for the analysis of rotationally symmetric scatterers using the MoM approach.

\begin{figure}
\centering
\includegraphics[width = 0.9 \hsize ]{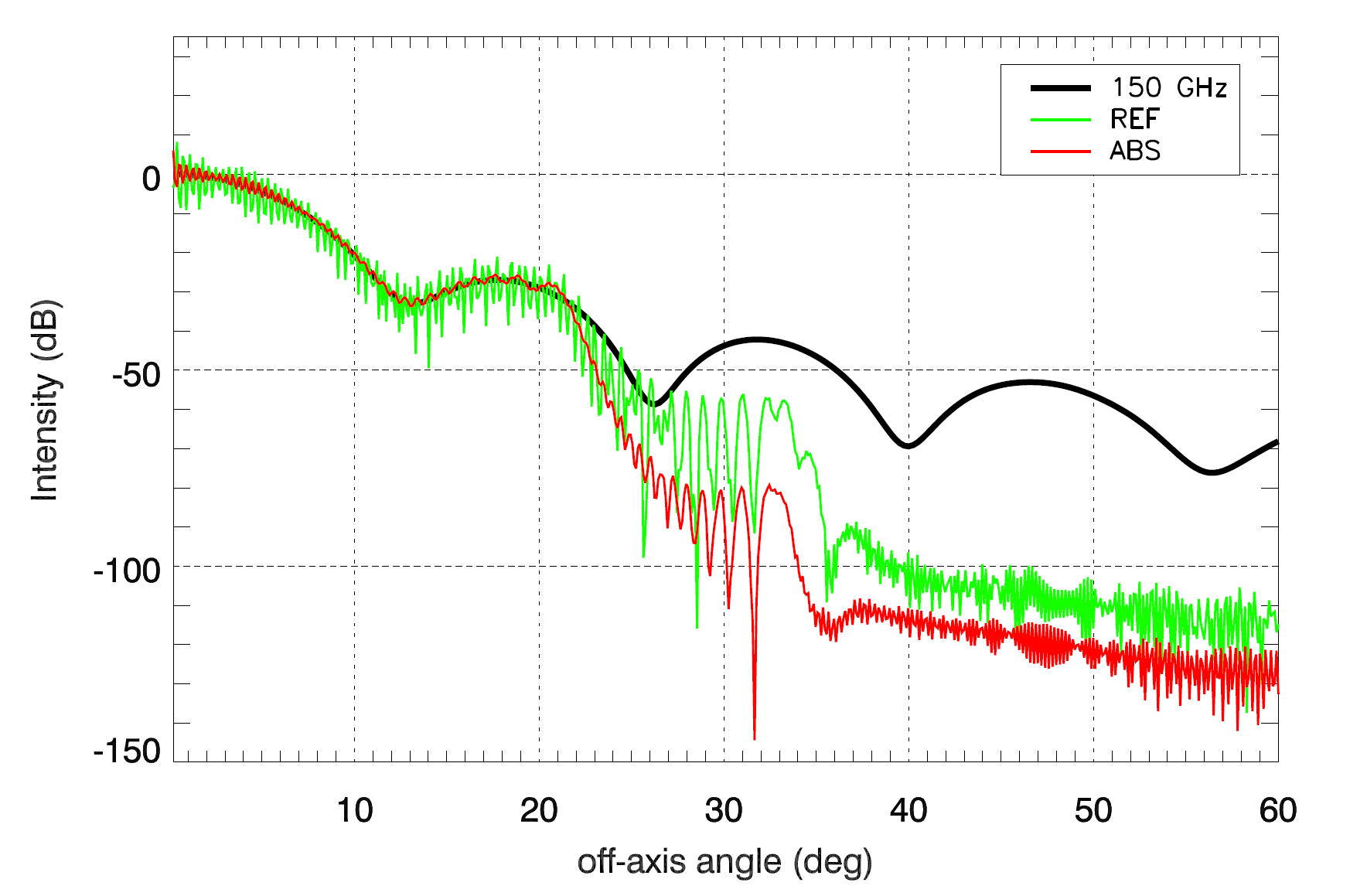}
\caption{Radial angular profile of the beam pattern with and without a conical FB of semi-aperture angle 14$^{\circ}$ and height 1~m. The different coloured traces refer to feed beam pattern (black line), reflective (green line) and absorptive (red line) internal surface solutions.}
\label{figFB_coating}
\end{figure}

For the absorptive solution we considered the possibility of covering the FB inner surface with 10~mm thick Eccosorb panels, a material with high lossy absorptive properties suitable for microwaves. We converged on ECCOSORB\circledR\ HR-10 with an outdoor fabric covering that is suitable for operating in harsh environments\footnote{Emerson \& Cuming, Microwave Products}. We assumed the following electrical parameters: an electric permittivity of 3.54, a magnetic permeability of 1 and a loss tangent of 0.057. The impact of both types of conical FB on the beam pattern of the central feedhorn at 150~GHz (figure~\ref{figbeam150}) is shown in figure~\ref{figFB_coating} for the case of a semi-aperture angle of 14$^{\circ}$ and a height of 1~m.

\begin{figure}
\centering
\includegraphics[width = 0.9 \hsize ]{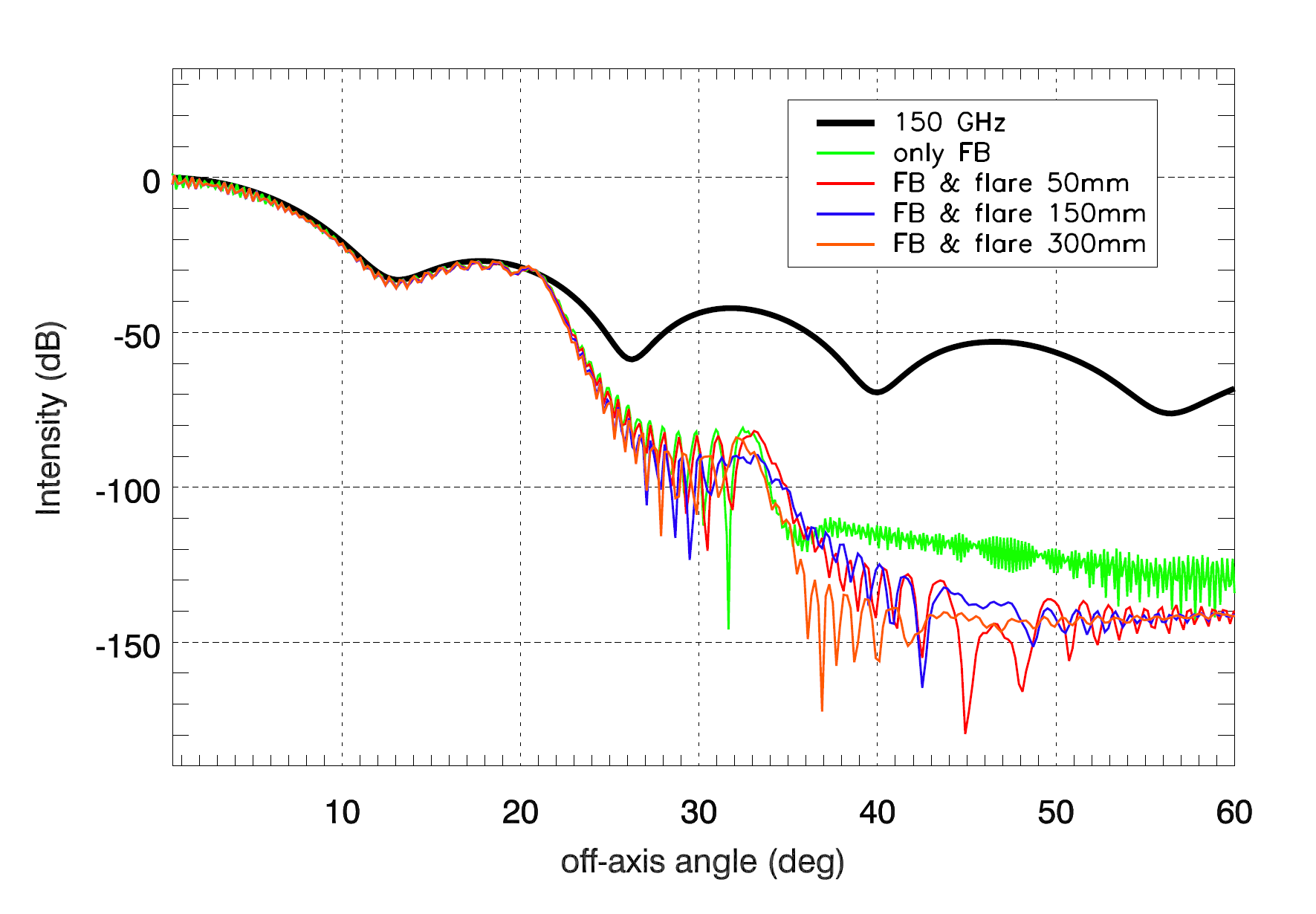}
\caption{Beam pattern cuts for the conical FB with an absorptive inner surface without and with a flare with three different radii of curvature. The coloured lines are defined in the legend.}
\label{figFB_flare}
\end{figure}

The presence of an absorptive inner surface on the FB increases the sidelobe rejection for angles larger than about 25$^{\circ}$ from boresight, compared with the nominal feedhorn beam pattern and the reflective solution. 
We added a flared edge to the entrance aperture of the FB, hereafter called the flare, with the aim of increasing the sidelobe rejection still further. Three values for the flare curvature radius, $R$ = 25$\lambda$, 75$\lambda$ and 150$\lambda$ (with $\lambda$ = 2 mm), were considered to check their impact on the final beam pattern. The insertion of the flare results in a further reduction of the sidelobes at angles larger than 35$^{\circ}$ from boresight, as shown in figure~\ref{figFB_flare}. The radius of curvature of the flare seems to have a small impact on decrease in sidelobes level. This allowed us to choose a flare with the smallest radius of curvature ($R$ = 50~mm), which makes mechanical fabrication easier.

We evaluated the additional 150~GHz radiative loading on a single detector, due to emission from the Eccosorb sheet on the inner surface of the FB, using the following equation:

\begin{equation}
P_{\mathrm{FB}} = A_{\mathrm{h}} t_{\mathrm{tot}} \epsilon_{\mathrm{opt}} \frac{ N_\mathrm{h} }{ N_{\mathrm{pix}}} \Delta \nu  B(\nu,T_{\mathrm{ec}}) \int_{\Delta \theta} \int_{\Delta \phi} B_{\mathrm{prim}}(\theta,\phi) \mathrm{sin} \theta \mathrm{co}s \theta d\theta d\phi
\label{eq:P_FB}
\end{equation}

\noindent
where $\theta$ and $\phi$ are the elevation and azimuthal angles with respect to the feedhorn beam centre, respectively, $\Delta \theta$ and $\Delta \phi$ are the angular ranges subtended by the FB at the central feed, $A_{\mathrm{h}}$ is the single feedhorn aperture area, $t_{\mathrm{tot}}$ is the final transmission of the whole optical chain, $\epsilon_{\mathrm{opt}}$ the optical efficiency, $N_\mathrm{h}$ is the number of feedhorns, $N_{\mathrm{pix}}$ is the total number of TESs in the focal plane, $\Delta \nu$ is the spectral bandwidth at 150~GHz, $B$ is the FB coating brightness for a temperature of $T_{\mathrm{ec}}$ and $B_{\mathrm{prim}}(\theta,\phi)$ is the single feedhorn beam pattern. The power collected by each pixel from the absorbing FB is of the order of 0.07~pW, well below the 0.33 pW we estimate as the contribution from the atmosphere.

\begin{figure}[hbtp]
\centering 
\includegraphics[width = 0.9 \hsize ]{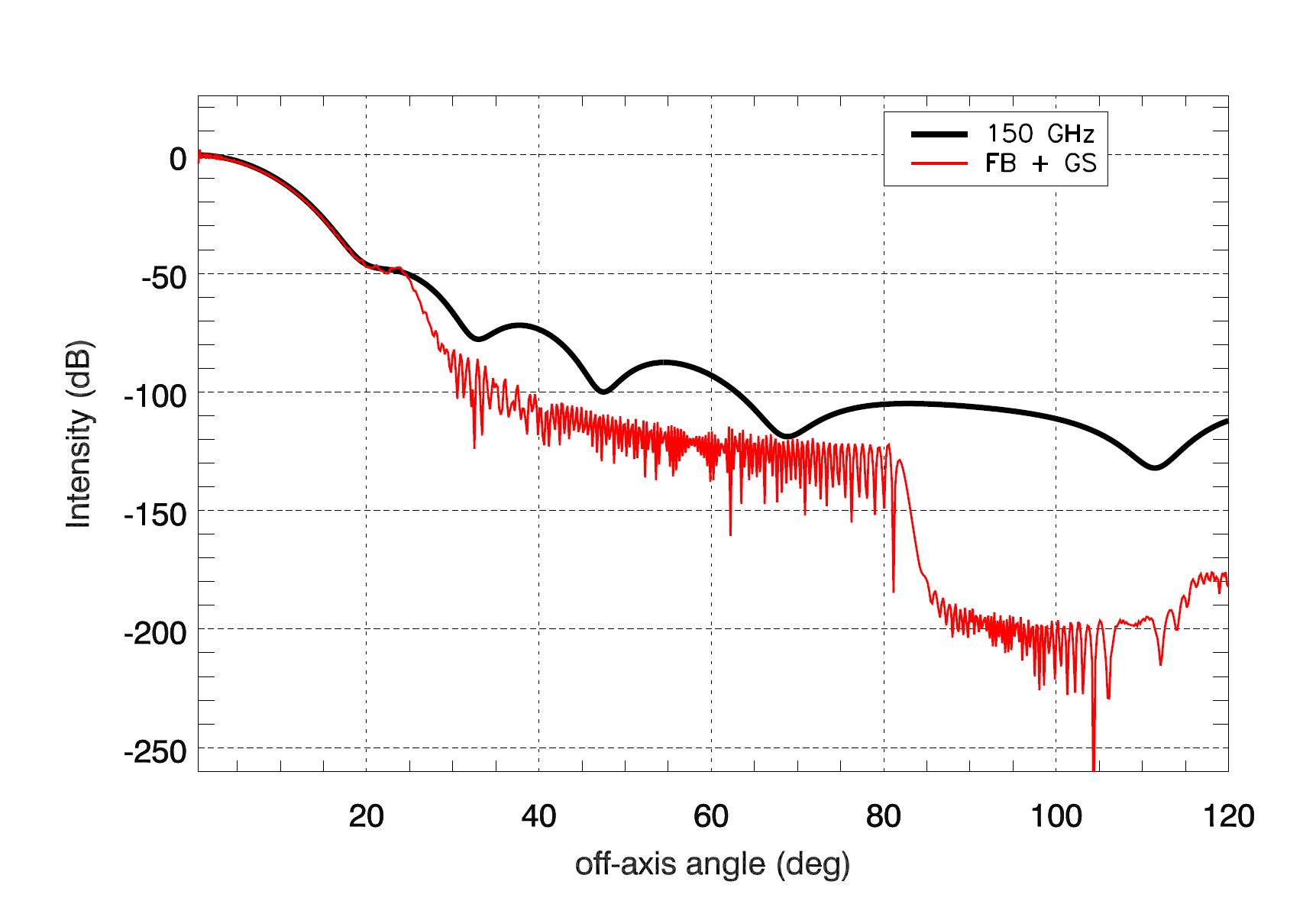}
\caption{Cut of the patterns of the central feed, modeled as a hybrid mode conical horn, (black line) and including the FB and GS (red line). The instrument is pointing to the zenith. The response drop at $\sim25^{\circ}$ is due to the presence of the FB; the impact of the GS edge is evident at $\sim80^{\circ}$. The small increase in the pattern approaching $120 ^{\circ}$ is due to radiation leakage between the base and the lateral surface of the GS.}
\label{figGS_beam}
\end{figure}

The whole instrument is moved using an altazimuth mount on the top of a well-adapted container which serves as a laboratory (figure~\ref{fig_lab_FB_GS}). It is surrounded by a GS in order to minimize the brightness contrast between the sky and the ground. Optical and mechanical constraints resulted in a GS with a conical shape and an aperture angle of $90^{\circ}$ ($45^{\circ}$ from the vertical), a base of 4~m in diameter, and a height of 1.5 m. The GS inner surface is reflective. Due to the large size, in terms of wavelength, of this scatterer, the following modelling results were obtained with a MultiGTD approach at 150~GHz only.
The impact of the FB plus the GS on the central feedhorn beam pattern is shown in figure~\ref{figGS_beam}, assuming the instrument is pointing toward the zenith. After the drop at $\sim25^{\circ}$ that is caused by the FB, a second knee in the beam pattern is evident at $\sim80^{\circ}$, owing to the presence of the GS edge. The increase in the feedhorn beam pattern approaching $120^{\circ}$ is due to radiation leakage between the base and the lateral surface of the GS. The simple model in \swname{GRASP} currently includes only the FB and GS and neglects other environment components. The far sidelobes in the beam patterns, therefore, have to be considered with caution, but at such a low level we don't expect this to be an issue. In the future, we plan to extend the full analysis by considering all the feedhorns and the final GS geometry.

\subsection{Design and manufacturing}{\label{sec:manuGSFB}}

\begin{figure}
\centering
\includegraphics[width = 1 \hsize ]{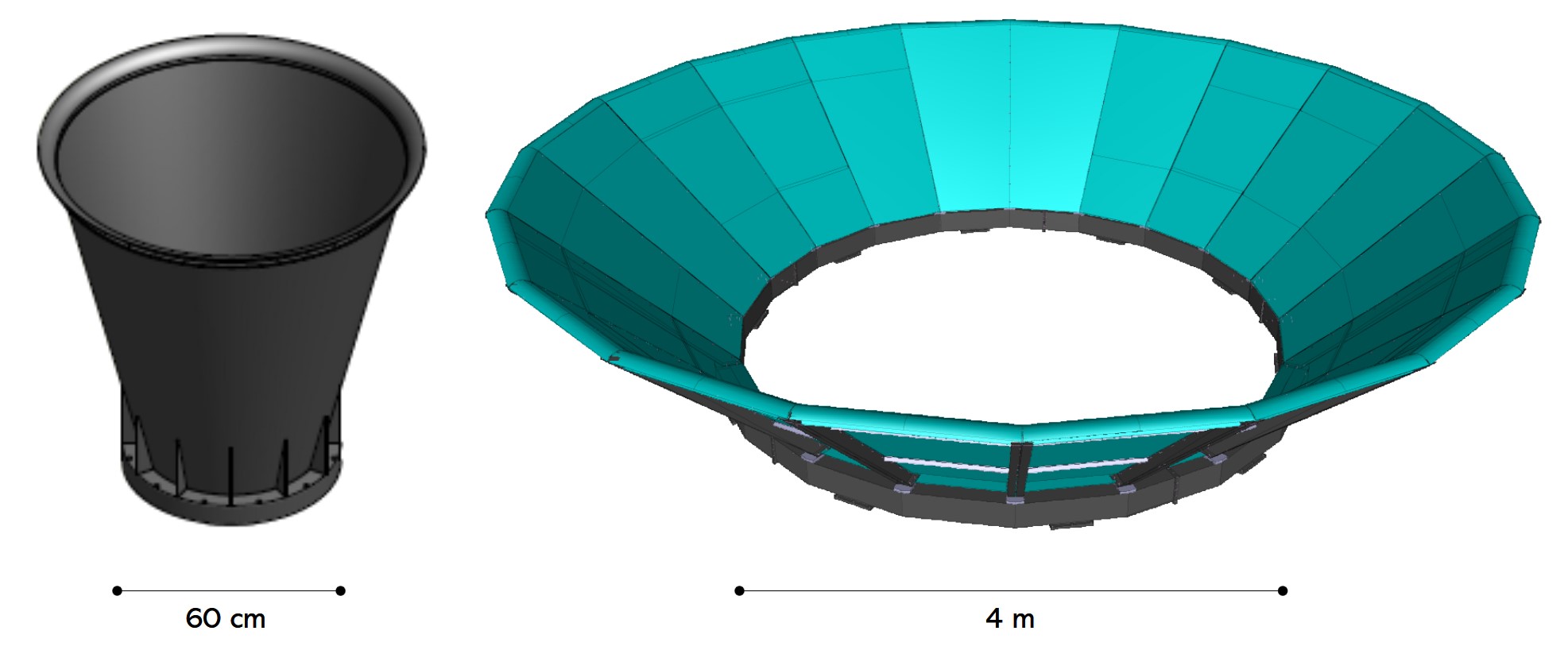}
\caption{Isometric views, not to scale, of the (left) FB and (right) GS.}
\label{figFB_GSdrawing}
\end{figure}

\begin{figure}
\centering
\includegraphics[width = 1 \hsize ]{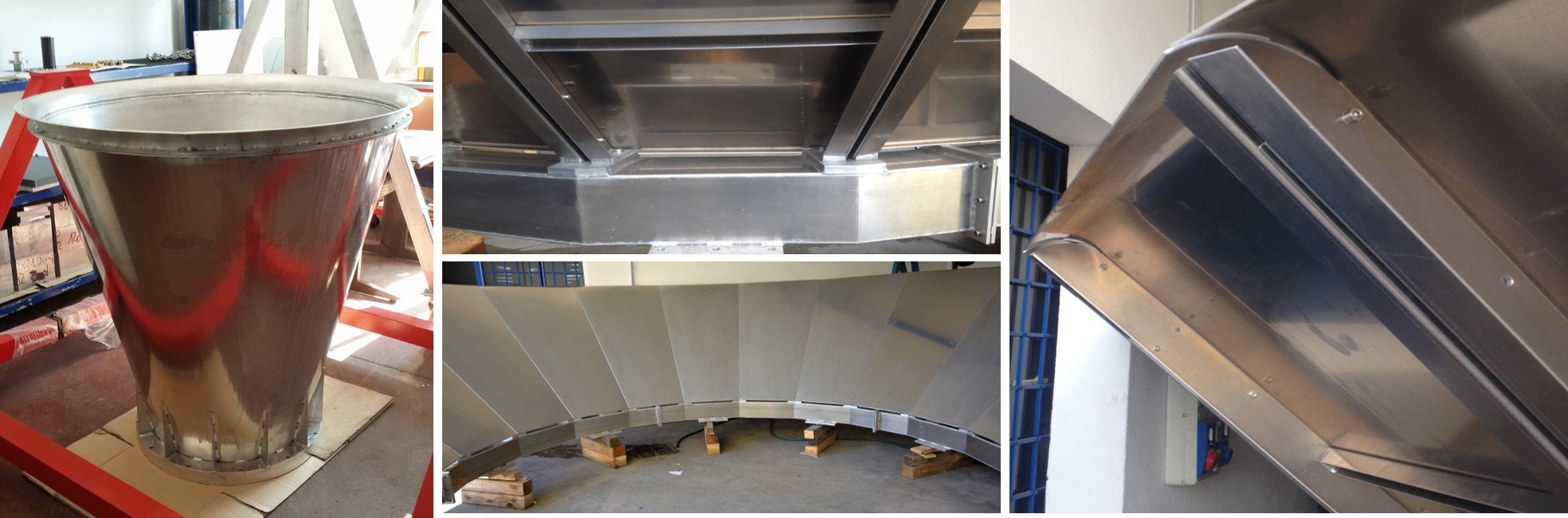}
\caption{The FB during manufacturing (left) and details of the GS: base (top middle), inner panels (bottom middle) and flare (right).}
\label{figFB_GS}
\end{figure}

Drawings of the FB and the GS are shown in figure~\ref{figFB_GSdrawing}. The FB was manufactured from a single panel of 2 mm thick aluminium alloy, formed into a truncated conical shape. The flare was added to the upper aperture, see figure~\ref{figFB_GS} (left). The inner surface was covered by panels of ECCOSORB\circledR\ HR-10. 

The GS is composed of 20 flat panels of 2-mm thick aluminium alloy, each one ending with a flare having a radius of curvature of 100~mm. The final conical shape has a semi-aperture of $45^{\circ}$ with an inner diameter equal to 4~m. In figure~\ref{figFB_GS} (right) a few details of the GS, before shipping to Argentina, are shown.

In order to be sure that environmental conditions on the site were satisfied, we performed a static structural analysis on the final GS design. We assumed that the shield is fixed on the top of the laboratory at 10 points distributed along the polygonal ring at the base. 
The load on the shield is mainly due to the pressure of the wind. We assumed a maximum velocity $v=38 \mbox{m}\mbox{s}^{-1}$, consistent with the {\sl Reglamento Argentino de la Acción del Viento Norma de aplicación: CIRSOC 102-2005}.
The mechanical analysis shows a low von Mises stress value and an acceptable total deformation of the panels, of less than 5 mm, well satisfying the requirements.

\section{Conclusions}
In this paper we have described the optics of the QUBIC CMB telescope, a unique instrument that uses the novel technique of bolometric interferometry to combine the sensitivity of an imager and the systematic control of an interferometer. We concentrated on the main components of the optical combiner and its shielding. The feedhorns were modelled using an electromagnetic mode-matching technique and their beams propagated through the combiner using physical optics. The combiner itself was designed to minimize truncations and aberrations.  Our analysis has shown that these produce a reduction in sensitivity, measured as the ratio between the real and ideal window function, to around 90\%. Because of the novel nature of the observing technique employed and the complex design of the instrument, a technical demonstrator was built to test the QUBIC design and its sub-systems. We describe the manufacturing and measurement of the mirrors and the procedure to align them. Initial beam measurements are in good agreement with prediction. A fixed groundshield and a forebaffle attached to the cryostat were designed to reduce contamination from ground and background sources. Physical optics modelling shows an increase in sidelobe rejection beyond $\sim25^{\circ}$ for the absorptive forebaffle and beyond $\sim80^{\circ}$ for the reflective groundshield.
The instrument is expected to be installed on the Alto Chorillo site in Argentina in 2021.

\acknowledgments

QUBIC is funded by the following agencies. France: ANR (Agence Nationale de la
Recherche) 2012 and 2014, DIM-ACAV (Domaine d’Intérêt Majeur-Astronomie et Conditions d’Apparition de la Vie), CNRS/IN2P3 (Centre national de la recherche scientifique/Institut national de physique nucléaire et de physique des particules), CNRS/INSU (Centre national de la recherche scientifique/Institut national et al de sciences de l’univers). Italy: CNR/PNRA (Consiglio Nazionale delle Ricerche/Programma Nazionale Ricerche in Antartide) until 2016, INFN (Istituto Nazionale di Fisica Nucleare) since 2017.  Argentina: MINCyT (Ministerio de Ciencia, Tecnología e Innovación), CNEA (Comisión Nacional de Energía Atómica), CONICET (Consejo Nacional de Investigaciones Científicas y Técnicas).
 
D. Burke and J.D. Murphy acknowledge funding from the Irish Research Council under the Government of Ireland Postgraduate Scholarship Scheme.  D. Gayer and S. Scully acknowledge funding from the National University of Ireland, Maynooth. D. Bennett acknowledges funding from Science Foundation Ireland.

\bibliographystyle{ieeetr}
\typeout{} 
\bibliography{qubic}

\end{document}